\documentclass[aps,prd,10pt,notitlepage,nofootinbib]{revtex4-1}

\usepackage[utf8]{inputenc}
\usepackage{amsmath,amssymb,amsfonts}
\usepackage{newtxtext,newtxmath}

\usepackage{bbold}
\usepackage{bm}
\usepackage{graphicx}
\usepackage[usenames,dvipsnames]{xcolor}
\usepackage{color}
\usepackage[colorlinks=true,linkcolor=blue,urlcolor=blue,citecolor=blue]{hyperref}
\usepackage{slashed}
\usepackage[english]{babel}
\usepackage{dcolumn}
\usepackage{pifont}
\usepackage{dsfont,mathrsfs}
\usepackage{cancel}
\usepackage{bigints}
\usepackage{accents}
\usepackage{soul}
\usepackage{multirow}

\usepackage{amsmath, amssymb}
\usepackage{bbold}
\usepackage{makecell}
\usepackage{simpler-wick}

\usepackage{orcidlink} %Added by Snigdha

\newcommand{\nn}{\nonumber}

\newcommand{\FB}[1]{\left(#1\right)}

\newcommand{\SB}[1]{\left\{#1\right\}}
\newcommand{\TB}[1]{\left[#1\right]}
\newcommand{\AB}[1]{\left<#1\right>}

\newcommand{\munu}{{\mu\nu}}

\newcommand{\IM}{\text{Im}}

\newcommand{\IMA}{\IM~\Pi_T}
\newcommand{\IMB}{\IM~\Pi_L}

\newcommand{\IMAU}{\IM~\Pi_T^{\rm Unitary}}
\newcommand{\IMAL}{\IM~\Pi_L^{\rm Landau}}

\newcommand{\IMBU}{\IM~\Pi_L^{\rm Unitary}}
\newcommand{\qL}{\sqrt{q_{\parallel}^2}}

\newcommand{\Pibar}{\overline{\Pi}}
\newcommand{\util}{\widetilde{u}}
\newcommand{\btil}{\widetilde{b}}
\newcommand{\Ntil}{\widetilde{N}}

\begin{document}
	%\begin{linenumbers}
	\title{Dilepton production from hot and magnetized hadronic matter}
	
	\author{Rajkumar Mondal$^{a,d}$}
	\email{rajkumarmondal.phy@gmail.com}
	
	\author{Nilanjan Chaudhuri\orcidlink{0000-0002-7776-3503}$^{a,d}$}
	\email{sovon.nilanjan@gmail.com}
	\email{n.chaudhri@vecc.gov.in}

	\author{Snigdha Ghosh\orcidlink{0000-0002-2496-2007}$^{b}$}
	\email{snigdha.physics@gmail.com}
	\email{snigdha@ggdckharagpur2.ac.in}
	\thanks{Corresponding Author}

	\author{Sourav Sarkar\orcidlink{0000-0002-2952-3767}$^{a,d}$}
	\email{sourav@vecc.gov.in}
	
	\author{ Pradip Roy$^{c,d}$}
	\email{pradipk.roy@saha.ac.in}

	\affiliation{$^a$Variable Energy Cyclotron Centre, 1/AF Bidhannagar, Kolkata - 700064, India}
	\affiliation{$^b$Government General Degree College Kharagpur-II, Paschim Medinipur - 721149, West Bengal, India}
	\affiliation{$^c$Saha Institute of Nuclear Physics, 1/AF Bidhannagar, Kolkata - 700064, India}
	\affiliation{$^d$Homi Bhabha National Institute, Training School Complex, Anushaktinagar, Mumbai - 400085, India}

	%+++++++++++++++++++++++++++++++++++++++++++++++++++++++++++++++++++++++++++++++++++++++++++++++++++++++++++++++++++++

\begin{abstract}
	The rate of dilepton emission from a magnetized hot hadronic medium is calculated in the framework of real time formalism of finite temperature field theory. We evaluate the one loop self-energy of neutral rho-meson containing thermo-magnetic propagators for the charged pions in the loop. The in-medium thermo-magnetic spectral function of rho obtained by solving the Dyson-Schwinger equation is shown to be proportional to the dilepton production rate. The study of the analytic structure of the neutral rho-meson spectral function in such a medium shows that in addition to the usual contribution coming from the Unitary cut beyond the two-pion threshold there is a non-trivial yield in the low invariant mass region originating due to the fact that the charged pions occupy different Landau levels before and after scattering with the neutral rho-meson and is purely a finite magnetic field effect. 
\end{abstract}

\maketitle

\section{Introduction}
The primary objective of modern Heavy Ion Collision (HIC) experiments at Large Hadron Collider (LHC) and Relativistic Heavy Ion Collider (RHIC) 
is to study hot and dense nuclear matter. The collision of two nuclei at ultra-relativistic energies leads to the liberation of the fundamental 
constituents of the nucleons forming a deconfined state of quarks and gluons in local thermal equilibrium. This form of the nuclear matter is 
known as the quark-gluon plasma (QGP), which, as suggested by the phenomenological studies, is the most perfect fluid created in 
nature~\cite{STAR:2005gfr,PHENIX:2004vcz,Romatschke:2007mq}. The fireball produced in HICs, cools via rapid expansion under its own pressure 
gradient going through various stages of evolution. However, the possibility of direct observation is strongly hindered as the QGP is very 
transient ($ \sim $ few fm/c). Thus to extract microscopic as well as bulk properties of QGP, one has to rely on indirect probes and observables 
such as spectra of electromagnetic probes (photon and dileptons), heavy quark production, quarkonia suppression, jet energy loss, collective 
flow, $ J/\psi $ suppression \textit{etc} (see Refs.~\cite{Wong:1995jf,Sarkar:2010zza,Florkowski:2010zz,Satz:2012zza} for a broad overview). 
Among these, electromagnetic probes~\cite{McLerran:1984ay,Kajantie:1986dh,Gale:1988vv,Weldon:1990iw,Ruuskanen:1990hx,Ruuskanen:1991au,Alam:1996fd,Alam:1999sc,Sarkar:2012ty}, 
owing to large mean free paths,  tend to leave the system without much interaction and, therefore are expected to  carry the information of the stage from where they are produced. This is the major 
advantage of the electromagnetic probes over hadrons which are emitted from the freeze out hyper-surface after undergoing rescattering.

The study of different $n$-point current-current correlation functions or  in-medium spectral functions of local currents is one of the primary 
theoretical tools to examine various properties of QGP. The electromagnetic spectral function is one such example which is obtained from the 
vector-vector current correlator which, in turn, is connected to the dilepton production rate (DPR) from the hot and dense 
medium~\cite{Gale:1988vv,Weldon:1990iw,Alam:1996fd,Alam:1999sc,Rapp:1999ej,Sarkar:2012ty}. In the QGP medium, the asymptotically free quarks can interact 
with an antiquark to produce a virtual photon, which decays into a dilepton. The emission rate resulting from these reactions has been 
extensively studied in Refs.~\cite{Kajantie:1986dh,Ruuskanen:1990hx,Ruuskanen:1991au}. However, there exist several other sources of 
dileptons (thermal and non-thermal) in HIC experiments~\cite{Ruuskanen:1991au,Wong:1995jf,Alam:1996fd,Cassing:1999es} which provide a substantial background. 
Among these, the contribution from the Drell-Yan process is well understood in the framework of perturbative quantum 
chromodynamics (QCD)~\cite{Craigie:1978bp,Grosso-Pilcher:1986iaq,CTEQ:1993hwr,Wong:1995jf,Alam:1996fd}. Dileptons can also be produced from the 
decays of hadron resonances, such as, $ \pi^0, \rho, \omega, J/\psi $, for which the yield can be estimated experimentally by invariant mass 
analysis~\cite{Wong:1995jf}.  However, the task to disentangle the photons and dilepton from the hadronic medium, produced after the 
phase transition/crossover, is a more daunting task. Therefore, a proper theoretical estimation of the photon and/or dilepton yield from hot 
and dense hadronic medium along with the possible modification of the hadronic properties below the critical temperature of the phase transition 
is of major importance to detect the electromagnetic signals from QGP. A significant amount of research has been carried out to evaluate the 
dilepton emission rate from hot and dense hadronic phase and it has been observed that the emission rate in the low invariant mass region 
is substantially modified~\cite{McLerran:1984ay,Weldon:1990iw,Gale:1990pn,Rapp:1999ej,Mallik:2016anp}.

Recent studies suggest that in non-central or asymmetric collisions of two heavy nuclei, very strong magnetic fields of the order $ \sim 10^{18} $ Gauss 
or larger might be generated due to the receding spectators~\cite{Kharzeev:2007jp,Skokov:2009qp}. The produced magnetic field decays very rapidly within few fm/c. 
However, it is found that both the QGP as well as hadronic medium possess finite electrical 
conductivity~\cite{Ding:2010ga,Amato:2013naa,Fernandez-Fraile:2005bew,Fernandez-Fraile:2009eug} which is expected to modify the decay process of this transient 
field according to relativistic magneto-hydrodynamics~\cite{Tuchin:2013apa,Tuchin:2015oka,Tuchin:2013ie,Gursoy:2014aka,Inghirami:2016iru,Inghirami:2019mkc}. 
Beside this, strong magnetic fields may also exist in several other physical systems. For example, in the interior of magnetars~\cite{Duncan:1992hi,Thompson:1993hn}, 
magnetic field $\sim 10^{15}$ Gauss can be present. Moreover, it is conjectured that in the early universe during the electroweak phase transition, 
magnetic fields as high as $ \sim 10^{23}$ Gauss might have been produced ~\cite{Vachaspati:1991nm,Campanelli:2013mea}. Since the strength of these magnetic 
fields is comparable to the typical QCD energy scale ($eB\sim \Lambda_\text{QCD}^2$), significant modifications can be found in various microscopic and bulk 
properties of the hadronic matter. For example, shear and bulk viscosity of magnetized hadronic matter using different methods has been 
studied in Refs.~\cite{Kadam:2014xka,Das:2019pqd,Dash:2020vxk,Ghosh:2022xtv}. Estimation of the electrical and the Hall conductivity of a hot and dense 
hadron gas in presence of uniform background field has been done in Refs.~\cite{Das:2019wjg,Kalikotay:2020snc}. In~Ref.\cite{Das:2020beh}, the magnetic field dependence of thermo-electric coefficients such as Seebeck, Nernst  \textit{etc} of a hadron gas are examined.

The modification of the DPR in the presence of a uniform background magnetic field from the QGP medium has been extensively studied in the literature using different 
approaches~\cite{Tuchin:2012mf,Tuchin:2013bda,Sadooghi:2016jyf,Bandyopadhyay:2016fyd,Bandyopadhyay:2017raf,Ghosh:2018xhh,Islam:2018sog,Ghosh:2020xwp,Hattori:2020htm,Chaudhuri:2021skc,Wang:2022jxx,Das:2021fma}. However, as the system cools down, it is expected that hadronic matter will be generated from QGP via a phase transition or crossover which has substantial contribution in the dilepton emission in the low invariant mass region. As discussed earlier, the presence of an external magnetic field leads to nontrivial modifications in transport properties of the hadronic matter. Hence it will be interesting to examine the effect of background magnetic field on the DPR from a hot and dense hadronic medium. An estimation of which is not readily available in the literature. The most important component in the calculation of DPR which determines the thresholds as well as the intensity of emission of dileptons is the imaginary part of the electromagnetic vector current correlator~\cite{Alam:1996fd,Alam:1999sc}. 
The latter quantity will be significantly modified owing the thermo-magnetic modification of the propagators of charged mesons. 
This will in turn modify the DPR from magnetized hadronic matter. 

In this work, we study the DPR from magnetized hot hadronic matter in terms of the spectral function of the neutral rho meson which is obtained from the electromagnetic vector current correlation function evaluated using the real time formalism (RTF) of Thermal Field Theory (TFT). The general formalism for the DPR is derived in the next section. In Section III the DPR is expressed in terms of the rho spectral function at finite temperature. This is extended to the case of non-zero magnetic field in Section IV making no approximations on the strength of the field. Section V contains the numerical results and we summarize in Section VI. Some details are provided in the appendix.

%~~~~~~~~~~~~~~~~~~~~~~~~~~~~~~~~~~~~~~~~~~~~~~~~~~~~~~~~~~~~~~~~~~~~~~~~~~~~~~~~~~~~~~~~~~~~~~~~~~~~~~~~~~~~~~~~~~~~~~~~~~~~~~~~~~~~~~~~~~
\section{FORMALISM}\label{Formalism}
The formalism to obtain the dilepton production rate (DPR) from a thermal system of hadrons has been discussed by many authors (see e.g. 
Refs.~\cite{Kajantie:1986dh, Gale:1987ki,Gale:1988vv, Ericson:1988gk, Alam:1999sc,Chanfray:1995jgo,Cleymans:1986na,Ruuskanen:1989tp,Mallik:2016anp}). 
Here we outline some essential steps following Ref.~\cite{Mallik:2016anp}. The emission rate of dileptons with four-momenta $q^\mu=(q^0,\bm{q})$ per 
unit space-time four-volume can be written as
\begin{equation}
	\frac{dN}{d^4xd^4q}=\frac{\alpha^2}{6\pi^3q^2}e^{-\beta q_0}L(q^2)\FB{-g_{\mu\nu}M^{+\mu\nu}}\label{DPR1}
\end{equation}
where $L(q^2)=\FB{1+\frac{2m_0^2}{q^2}}\sqrt{1-\frac{4m_0^2}{q^2}}$, $m_0$ is the leptonic mass and $M^{+\mu\nu}$ is the Fourier 
transform of two point vector current correlator
\begin{equation}
	M^{+\mu\nu}=\int\! d^4x e^{iq.x}\AB{J_h^\mu(x)J_h^\nu(0)} \label{eq.M+}
\end{equation}
in which $\AB{\cdots}$ denotes thermal ensemble average and $J_h^\mu(x)$ is the electromagnetic current of hadrons. 
The quantity $M^{+\mu\nu}$ can be calculated using standard techniques of finite temperature field theory as follows.

In the RTF of TFT, the two-point correlation functions assume a $2\times2$ matrix structure on account of the shape of the contour in the complex time plane ~\cite{Bellac:2011kqa,Mallik:2016anp}. We start with the Fourier transform of the \emph{time-ordered} two point function
\begin{equation}
	M_{ab}^{\mu\nu}=i\int\! d^4x~e^{iq.x}\AB{T_c\SB{J_h^\mu\FB{x}J_h^\nu\FB{0}}}_{ab}\label{Mab1}
\end{equation}
where $T_c$ indicates time-ordering with respect to the time contour $c$. The thermal indices $a,b\in\SB{1,2}$  correspond to the fact that the two points can be 
chosen on either of the two horizontal segments of the contour $c$. The quantity in Eq.~\eqref{Mab1} can be expressed in diagonal form as
\begin{equation}
	M^{\mu\nu}=U \begin{pmatrix} \overline{M}^{\mu\nu} & 0 \\ 0 & -\overline{M}^{*\mu\nu} \end{pmatrix} U\label{Mbar}
\end{equation}
by means of the matrix $U=\begin{pmatrix} \sqrt{n+1} & \sqrt{n} \\ \sqrt{n} & \sqrt{n+1} \end{pmatrix}$ where $n=\frac{1}{e^{\beta|q_0|}-1}$ is a 
thermal distribution like function in which $\beta=1/T$ is the inverse temperature. 
The diagonal element $\overline{M}^{\mu\nu}$ appearing on the RHS of Eq.~\eqref{Mbar} is an analytic function and is obtainable from any one of the components of $M_{ab}^{\mu\nu}$. 
It is related, say for example to the 11-component as
\begin{equation}
	\text{Re}~\overline{M}^{\mu\nu}(q)=\text{Re}~{M}_{11}^{\mu\nu}(q)~;~\text{Im}~\overline{M}^{\mu\nu}(q)=\tanh\FB{\frac{|q_0|}{2T}}~\text{Im}~{M}_{11}^{\mu\nu}(q) \label{ReImM}.
\end{equation}
Now, using a spectral representation~\cite{Mallik:2016anp}, one can relate the quantity $M^{+\mu\nu}$ appearing in Eq.~\eqref{eq.M+} with the imaginary part of the analytic function $\overline{M}^{\mu\nu}$ as
\begin{equation}
	M^{+\mu\nu}\FB{q}=\frac{2e^{\beta q_0}}{e^{\beta q_0}-1}~\text{Im}~\overline{M}^{\mu\nu}\FB{q}=\epsilon(q^0)\frac{2e^{\beta q_0}}{e^{\beta q_0}+1}~\text{Im}~M^{\mu\nu}_{11}\FB{q}\label{Mplus1}
\end{equation}
where $\epsilon(q^0)$ is the sign function. 
Substituting Eq.~\eqref{Mplus1} into Eq.~\eqref{DPR1}, we get the DPR in terms of $\overline{M}^{\mu\nu}$ as
\begin{equation}
	\frac{dN}{d^4xd^4q}=\frac{\alpha^2}{3\pi^3q^2}\frac{1}{e^{\beta q_0}-1}L(q^2)\FB{-g_{\mu\nu}~\text{Im}~\overline{M}^{\mu\nu}\FB{q}}.
	\label{DPR1.1}
\end{equation}

In order to calculate $M^{\mu\nu}_{11}$, we now require the explicit form of the hadronic local vector current $J_h^\mu(x)$. 
Considering only the iso-vector rho-mesons in VDM~\cite{Sarkar:2000ag,Sarkar:2012ty,Mallik:2016anp, Bhaduri:1988gc, Ericson:1988gk}, the hadronic current can be expressed as
\begin{equation}
	J_h^\mu(x)=J_{\FB{\rho}}^\mu(x)=F_\rho m_\rho\rho^\mu(x) \label{eq.Jh}
\end{equation}
where $\rho^\mu(x)$ is the Heisenberg field corresponding to the $\rho^0$ meson and the coupling $F_\rho$=156 MeV is obtained from the 
decay rate $\Gamma_{\rho^0\to e^{+}e^{-}}=7.0$ keV~\cite{Mallik:2016anp}. Substituting Eq.~\eqref{eq.Jh} into Eq.~\eqref{Mab1}, and applying the Wick's theorem, we arrive at
\begin{equation}
	\text{Im}~M^{\mu\nu}_{11}(q)=F_\rho^2m_\rho^2~\text{Im}~D^{\mu\nu}_{11}(q) \label{eq.M11}
\end{equation}
where $D^{\mu\nu}_{11}\FB{q}$ is the 11-component of the exact thermal propagator of the $\rho^0$-meson. 
Making use of Eqs.~\eqref{eq.M11} and \eqref{Mplus1}, Eq.~\eqref{DPR1.1} can be written as,
\begin{equation}
	\frac{dN}{d^4xd^4q}=\frac{\alpha^2}{\pi^3q^2}\frac{1}{e^{\beta q_0}-1}L(q^2)F_\rho^2m_\rho^2\FB{-\frac{1}{3}g_{\mu\nu}~\text{Im}~\overline{D}^{\mu\nu}\FB{q}}.
	\label{DPR1.2}
\end{equation}
where, $\overline{D}^{\mu\nu}$ is the diagonal element corresponding to the real time interacting $\rho^0$ propagator and is related to $ D^\munu_{11} (q)$ by means of relation analogous to Eq.~\eqref{ReImM}. 
The term within the large parenthesis on the RHS of Eq.~\eqref{DPR1.2} can be identified as the in-medium spectral function of the $\rho^0$ meson i.e. 
$\mathcal{A}(q;T) = -\frac{1}{3}g^{\mu\nu}~\text{Im}~\overline{D}_{\mu\nu}\FB{q}$. Thus in terms of spectral function, the DPR in Eq.~\eqref{DPR1.2} can be written as
\begin{equation}
	\frac{dN}{d^4xd^4q}=\frac{\alpha^2}{\pi^3q^2}f_\text{BE}(q_0)L(q^2)F_\rho^2m_\rho^2\mathcal{A}(q;T).
	\label{DPR2}
\end{equation}
where $f_\text{BE}(x)=\FB{e^{x/T}-1}^{-1}$ is the Bose-Einstein thermal distribution function. 
Our next task is to calculate the $\rho^0$-meson spectral function $\mathcal{A}(q;T)$ in a thermal medium in the presence of external magnetic field. 
For the sake of completeness and continuity, we will first calculate $\mathcal{A}(q;T)$ in absence of magnetic field in the next section. 
Later in Sec.~\ref{DPRB}, we will evaluate $\mathcal{A}(q;T,eB)$ for a general thermo-magnetic background.

%~~~~~~~~~~~~~~~~~~~~~~~~~~~~~~~~~~~~~~~~~~~~~~~~~~~~~~~~~~~~~~~~~~~~~~~~~~~~~~~~~~~~~~~~~~~~~~~~~~~~~~~~~~~~~~~~~~~~~~~~~~~~~~~~~~~~
\section{RHO SPECTRAL FUNCTION AND DPR IN ABSENCE OF MAGNETIC FIELD}\label{DRPB0}
As discussed in the last section, the essential quantity in the DPR which contains the dynamics of the hadronic medium is (imaginary part of) the 
exact rho meson propagator or the in-medium spectral function. The diagonal component of the real time exact $\rho^0$-propagator $\overline{D}^{\mu\nu}$ can be obtained 
in terms of the bare $\rho^0$-propagator $\overline{D}_{(0)}^{\mu\nu}$ and the analytic thermal self-energy function $\overline\Pi_{\alpha\beta}$ by solving the following Dyson-Schwinger equation~\cite{Mallik:2016anp,Bellac:2011kqa}:
\begin{equation}
	\overline D^{\mu\nu}=\overline D^{\mu\nu}_{(0)}-\overline D^{\mu\alpha}_{(0)}\overline\Pi_{\alpha\beta} \overline D^{\beta\nu}\label{Dyson}
\end{equation}
where
\begin{equation}
	\overline D^{\mu\nu}_{(0)}(q) = \left(-g^{\mu\nu}+\frac{q^\mu q^\nu}{m_\rho^2}\right)\frac{-1}{q^2-m_\rho^2+i\epsilon}\label{rhoprop}.
\end{equation}
The analytic thermal self-energy function $\overline\Pi^{\alpha\beta}$ (which is the diagonal element of $U^{-1}\Pi^{\alpha\beta}U^{-1}$ in thermal space), can be obtained from 
the 11-component $\Pi^{\alpha\beta}_{11}$ by means of relations analogous to Eq.~(\ref{ReImM}). The latter is now evaluated in perturbation theory 
using the effective field theoretic Lagrangian~\cite{Krehl:1999km}
\begin{equation}
	\mathcal{L}_\text{int}=-g_{\rho\pi\pi}~(\partial_\mu\bm{\rho}_\nu)\cdot\FB{\partial^\mu\bm{\pi}\times\partial^\nu\bm{\pi}} \label{Lint} 
\end{equation}
where, $\bm{\rho}_\mu$ and $\bm{\pi}$ are the iso-vector fields corresponding to the rho-mesons and pions respectively, and, $g_{\rho\pi\pi}=20.72$ GeV$^{-2}$ is the value 
of the coupling constant determined from the  decay width $\Gamma_{\rho\to\pi\pi}=155.8$ MeV.  

\begin{figure}[h]
	\includegraphics[scale=.5]{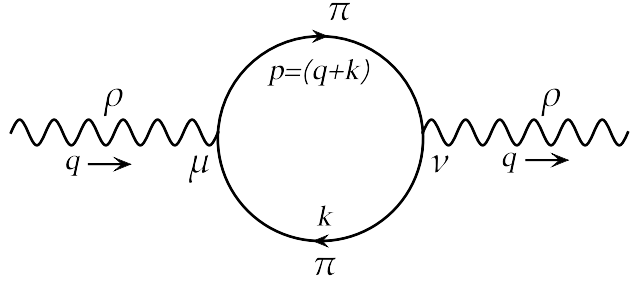}	
	\caption{Feynman diagram for one-loop self-energy of $\rho$-meson.}\label{Fig}
\end{figure}

\begin{figure}[h]
	\includegraphics[scale=1.2]{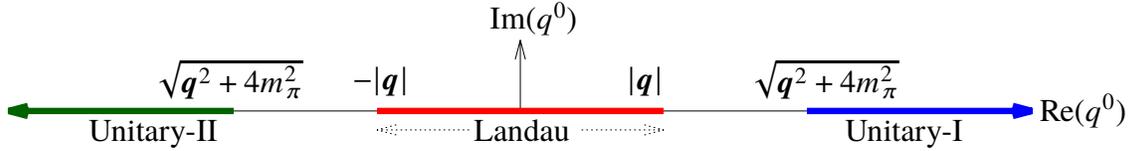}
	\caption{Analytic structure of $\Pibar^{\mu\nu}(q^0,\bm{q})$ in complex plane of $q^0$ for a given $\bm{q}$. Unitary-I (denoted by blue line) corresponds to the domain of physical dilepton production.}\label{KDM1}
\end{figure}

Using Eq.~\eqref{Lint}, the expression for the 11-component of the one-loop self-energy matrix of $\rho^0$ obtained by applying Feynman rules to the graph shown in Fig.~\ref{Fig} is given by
\begin{equation}
	\Pi^{\mu\nu}_{11}(q;T)=i \int\! \frac{d^4k}{(2\pi)^4}N^{\mu\nu}(q,k)D_{11}(k)D_{11}(p=q+k)\label{DeltaPi7}
\end{equation}
where
\begin{equation}
	N^{\mu\nu}(q,k)=g^2_{\rho\pi\pi}\TB{q^4k^\mu k^\nu+(q\cdot k)^2q^\mu q^\nu-q^2(q\cdot k)\FB{q^\mu k^\nu+q^\nu k^\mu}} \label{eq.N0}
\end{equation}
contains the factors coming from the interaction vertices and $D_{11}(k)$ is the 11-component of the real time thermal pion propagator expressed as~\cite{Mallik:2016anp,Bellac:2011kqa}
\begin{equation}
	D_{11}(k)=\frac{-1}{k^2-m_\pi^2+i\epsilon}+2\pi i \eta(k\cdot u)\delta(k^2-m_\pi^2)\label{D111}.
\end{equation}
in which $\eta(x)=\Theta(x)f_\text{BE}(x)+\Theta(-x)f_\text{BE}(-x)$, $u^\mu$ is the medium four-velocity and $m_\pi$ is pion rest mass.
In the Local Rest Frame (LRF) of the medium, $u^\mu_\text{LRF}\equiv(1,\bm{0})$.
As mentioned earlier, the analytic thermal self-energy function $\Pibar^{\mu\nu}(q)$ can be obtained from $\Pi^{\mu\nu}_{11}(q)$ using the relations
\begin{eqnarray}
	\text{Re}~\overline{\Pi}^{\mu\nu}(q^0,\bm{q})=\text{Re}~\Pi^{\mu\nu}_{11}(q^0,\bm{q}) ; ~~
	\text{Im}~\overline{\Pi}^{\mu\nu}(q^0,\bm{q})=\tanh\FB{\frac{|q^0|}{2T}}\text{Im}~\Pi^{\mu\nu}_{11}(q^0,\bm{q})\label{ImDeltaPi1}.
\end{eqnarray}
On substituting Eq.~\eqref{D111} into Eq.~\eqref{DeltaPi7} and performing the $dk^0$ integration  
we get the real and imaginary parts of $\rho^0$-meson thermal self-energy function using Eq.~\eqref{ImDeltaPi1} as
\begin{eqnarray}
	\text{Re}~\Pibar^{\mu\nu}(q^0, \bm{q};T) &=& \text{Re}~\Pi^{\mu\nu}_\text{Pure-Vac}(q)
	+\int\!\frac{d^3k}{(2\pi)^3}~\mathcal{P} \bigg[ \frac{f_k}{2\omega_k} \bigg\{ \frac{N^{\mu\nu}(k^0=-\omega_k)}{(q^0-\omega_k)^2-(\omega_p)^2} 
	+\frac{N^{\mu\nu}(k^0=\omega_k)}{(q^0+\omega_k)^2-(\omega_p)^2} \bigg\}  \nn \\ 
	&& 	+\frac{f_p}{2\omega_p} \bigg\{ \frac{N^{\mu\nu}(k^0=-q^0-\omega_p)}{(q^0+\omega_p)^2-(\omega_k)^2}
	+\frac{N^{\mu\nu}(k^0=-q^0+\omega_p)}{(q^0-\omega_p)^2-(\omega_k)^2} \bigg\} \bigg], \label{eq.repi.0}
	\\
	%\end{eqnarray}
	%\begin{eqnarray}
	\text{Im}~\Pibar^{\mu\nu}(q^0, \bm{q};T)&=&-\tanh\FB{\frac{|q^0|}{2T}}\pi\int\!\frac{d^3k}{(2\pi)^3}\frac{1}{4\omega_k\omega_p} \nn \\
	&& \times \TB{(1+f_k+f_p+2f_kf_p) \SB{N^{\mu\nu}(k^0=-\omega_k)\delta(q^0-\omega_k-\omega_p)+N^{\mu\nu}(k^0=\omega_k)\delta(q^0+\omega_k+\omega_p)} \right. \nn \\ 
		&& \left. + (f_k+f_p+2f_kf_p) \SB{N^{\mu\nu}(k^0=-\omega_k)\delta(q^0-\omega_k+\omega_p) + N^{\mu\nu}(k^0=\omega_k)\delta(q^0+\omega_k-\omega_p)}
	}\label{ImPi2}
\end{eqnarray}
where, $\omega_k=\sqrt{m_\pi^2+\bm{k}^2}$, $\omega_p=\sqrt{m_\pi^2+\bm{p}^2}=\sqrt{m_\pi^2+(\bm{q}+\bm{k})^2}$, $f_k = f_\text{BE}(\omega_k)$, $f_p = f_\text{BE}(\omega_p)$ 
and $\mathcal{P}$ denotes the Cauchy principal value integration. In Eq.~\eqref{eq.repi.0}, the quantity $\text{Re}~\Pi^{\mu\nu}_\text{Pure-Vac}(q)$ is given by
\begin{equation}
	\Pi^{\mu\nu}_\text{Pure-Vac}(q)=i \int\! \frac{d^4k}{(2\pi)^4}\frac{N^{\mu\nu}(q,k)}{(k^2-m_\pi^2+i\epsilon)\{(q+k)^2-m_\pi^2+i\epsilon\}} \label{eq.pure.vac}
\end{equation}
which is the temperature independent pure vacuum contribution to the self-energy. We also note that one of the integrations $d(\cos\theta)$ of Eq.~\eqref{ImPi2} can 
be analytically performed using the Dirac delta functions present in the integrand. The arguments of delta functions in Eq.~\eqref{ImPi2} correspond to energy-momentum conservation and they are non-vanishing in certain domains of energy ($q^0$) for a given three momentum $\bm{q}$. 
They are responsible for producing branch cuts of the self-energy function in the complex $q^0$ plane. The branch cuts due to the four Dirac delta functions in Eq.~\eqref{ImPi2} are termed as Unitary-I, Unitary-II, Landau-II and Landau-I cuts respectively as they appear in the equation. The non-vanishing kinematic domains for the Unitary-I and II are $\sqrt{\bm{q}^2+4m_\pi^2}<q^0<\infty$ and $-\infty<q^0<-\sqrt{\bm{q}^2+4m_\pi^2}$, respectively. For the two Landau cuts, the same is given by $|q^0|<|\bm{q}|$. The cut structure of the self-energy function is shown in Fig.~\ref{KDM1}. The cuts represent different physical processes such as decay or scattering. The Unitary-I cut corresponds to the decay $\rho^0\to\pi^+\pi^-$ (and the time reversed process) and the Landau cuts indicate the scattering of $\rho^0$ off pions in the medium. In the physical time like region (defined by $q^0>0$ and $q^2>0$), only the Unitary-I cut contributes. It may be noted that if the loop particles were of different masses, a non-trivial Landau cut would have appeared in the physical time like region. The kinematic domain for such non-trivial Landau cut would be $|\bm{q}|<q^0<\sqrt{\bm{q}^2+\Delta m^2}$ where $\Delta m$ is the mass difference between the two loop particles (in our case $\Delta m=0$). 

Having obtained the self-energy, we now proceed to obtain the exact propagator by solving Eq.~\eqref{Dyson}. 
It is convenient to decompose the self-energy tensor into independent covariants as~\cite{Ghosh:2019fet}
\begin{equation}
	\Pibar^{\mu\nu} (T)=\Pi_T P_T^{\mu\nu}+\Pi_L P_L^{\mu\nu}\label{LorPi}
\end{equation}
where
\begin{equation}
	P_T^{\mu\nu}=\FB{g^{\mu\nu}-\frac{q^\mu q^{\nu}}{q^2}-\frac{\util^\mu\util^\nu}{\util^2}}~~;~~P_L^{\mu\nu}=\frac{\util^\mu\util^\nu}{\util^2}
	\label{eq.proj.1}
\end{equation}
in which $\util^\mu=u^\mu-\FB{\frac{q\cdot u}{q^2}}q^\mu$ is a vector orthogonal to $q^\mu$. 
The form factors $\Pi_T$ and $\Pi_L$ appearing on the RHS of Eq.~\eqref{LorPi} comes out to be
\begin{equation}
	\Pi_T=\frac{1}{2}\FB{g_\munu\Pi^\munu-\frac{1}{\util^2}u_\mu u_\nu\Pi^{\mu\nu}}~~;~~\Pi_L=\frac{1}{\util^2}u_\mu u_\nu\Pi^{\mu\nu}.
\end{equation}
Using Eqs.~\eqref{rhoprop} and \eqref{LorPi}, Eq.~(\ref{Dyson}) is solved to get the interacting $\rho^0$-meson propagator as
\begin{equation}
	\overline{D}^{\mu\nu}(T) =\frac{P^{\mu\nu}_T}{q^2-m_\rho^2+\Pi_T}+\frac{P^{\mu\nu}_L}{q^2-m_\rho^2+\Pi_L}-\frac{q^\mu q^\nu}{q^2m_\rho^2}
\end{equation}
whose imaginary part gives the in-medium spectral function of rho meson as
\begin{equation}
	\mathcal{A}(q;T)=-\frac{1}{3}g^{\mu\nu}\text{Im}~\overline{D}_{\mu\nu}=\frac{1}{3}\TB{\frac{2\text{Im}\Pi_T}{(q^2-m_\rho^2+\text{Re}\Pi_T)^2+(\text{Im}\Pi_T)^2}
		+\frac{\text{Im}\Pi_L}{(q^2-m_\rho^2+\text{Re}\Pi_L)^2+(\text{Im}\Pi_L)^2}}. \label{eq.spec.1}
\end{equation}
Having obtained the spectral function $\mathcal{A}(q;T)$, it is now straight forward to calculate the DPR by substituting Eq.~\eqref{eq.spec.1} into Eq.~\eqref{DPR2}.
We note that the kinematic domain for dilepton production is shown in Fig.~\ref{KDM1} by the blue line (Unitary-I cut) where the spectral function is non-zero; 
and there is no contribution to dilepton production from the Landau cuts for physical dileptons having $q^0>0$ and $q^2>0$. For comparison, the DPR from the hadronic matter commonly obtained in the literature (for example in Refs.~\cite{Gale:1988vv,Gale:1990pn} by C. Gale and J. Kapusta) is provided in Appendix~\ref{A1}.

%~~~~~~~~~~~~~~~~~~~~~~~~~~~~~~~~~~~~~~~~~~~~~~~~~~~~~~~~~~~~~~~~~~~~~~~~~~~~~~~~~~~~~~~~~~~~~~~~~~~~~~~~~~
\section{RHO SPECTRAL FUNCTION AND DPR IN PRESENCE OF MAGNETIC FIELD}\label{DPRB}
Let us now consider a constant background magnetic field $B$ along the positive $\hat{\bm{z}}$ direction in addition to finite temperature.
In such a thermo-magnetic background, the 11-component of one-loop self-energy of neutral rho meson in Eq.~\eqref{DeltaPi7} modifies to
\begin{equation}
	\Pibar^{\mu\nu}_{11}(q;T,eB)=i \int\! \frac{d^4k}{(2\pi)^4}N^{\mu\nu}(q,k)D^B_{11}(k)D^B_{11}(p=q+k)
	\label{eq.Pi.B.1}
\end{equation}
where, $e$ is the electronic charge of a proton, $D^B_{11}(k)$ is the 11-component of the real time charged pion propagator in Schwinger representation given by~\cite{Ghosh:2019fet,Ayala:2016awt}
\begin{equation}
	D^B_{11}(k)=\sum_{l=0}^{\infty}2(-1)^le^{-\alpha_k}L_l(2\alpha_k)\TB{\frac{-1}{k_\parallel^2-m^2_l+i\epsilon}+2\pi i\eta(k\cdot u)\delta(k_\parallel^2-m^2_l)}\label{DeltaPi2}.
\end{equation}
Here $l$ is the Landau level index, $\alpha_k=-k_\perp^2/eB$, $L_l(z)$ is Laguerre polynomial of order $l$, $m_l=\sqrt{m^2_\pi+(2l+1)eB}$ is the effective Landau level dependent pion mass, $k_{\parallel,\perp}^\mu = g_{\parallel,\perp}^\munu k_\nu$ with $g_\parallel^\munu=\text{diag}(1,0,0,-1)$ and $g_\perp^\munu=\text{diag}(0,-1,-1,0)$. 
Note that in this convention $k_\parallel^2=(k_0^2-k_z^2)$ and $k_\perp^2=-(k_x^2+k_y^2)<0$.

	Now, substituting Eq.~\eqref{DeltaPi2} into \eqref{eq.Pi.B.1} and performing the $dk^0$ integration and using Eq.~\eqref{ImDeltaPi1}, we get the real and imaginary parts of the $\rho^0$ self-energy as	
	\begin{eqnarray}
	\text{Re}~{\Pibar}^{\mu\nu}(q^0,\bm{q};T,eB)&=&\text{Re}~\Pibar^{\mu\nu}_{\rm~Vac}(q,eB)+\sum_{n=0}^{\infty}\sum_{l=0}^{\infty}\int\frac{d^3k}{\FB{2\pi}^3}~\mathcal{P}\TB{\frac{f_k^l}{2\omega^l_k}\SB{\frac{\Ntil^{\mu\nu}_{nl}(k^0=-\omega^l_k)}{(q^0-\omega^l_k)^2-(\omega^n_p)^2}+\frac{\Ntil^{\mu\nu}_{nl}(k^0=\omega^l_k)}{(q^0+\omega^l_k)^2-(\omega^n_p)^2}}\nn\right.\\&&\left.+\frac{f_p^n}{2\omega^n_p}\SB{\frac{\Ntil^{\mu\nu}_{nl}(k^0=-q^0-\omega^n_p)}{(q^0+\omega^n_p)^2-(\omega^l_k)^2}+\frac{\Ntil^{\mu\nu}_{nl}(k^0=-q^0+\omega^n_p)}{(q^0-\omega^n_p)^2-(\omega^l_k)^2}}}\label{eq.repi.B.d3k},
%	\end{eqnarray}
	\\
%	\begin{eqnarray}
	\text{Im}~{\Pibar}^{\mu\nu}(q^0,\bm{q};T,eB)&=&-\tanh\FB{\frac{|q^0|}{2T}}\sum_{n=0}^{\infty}\sum_{l=0}^{\infty}~\pi\int\frac{d^3{k}}{\FB{2\pi}^3}~\frac{1}{4\omega^l_k\omega^n_p}\left[\SB{1+f^l_k+f^n_p+2f^l_kf^n_p}\nn\right.\\&&\left.\times\SB{\Ntil^{\mu\nu}_{nl}(k^0=-\omega^l_k)\delta(q^0-\omega^l_k-\omega^n_p)+\Ntil^{\mu\nu}_{nl}(k^0=\omega^l_k)\delta(q^0+\omega^l_k+\omega^n_p)}+\SB{f^l_k+f^n_p+2f^l_kf^n_p}\nn\right.\\&&\left.\times\SB{\Ntil^{\mu\nu}_{nl}(k^0=-\omega^l_k)\delta(q^0-\omega^l_k+\omega^n_p)+\Ntil^{\mu\nu}_{n,l}(k^0=\omega^l_k)\delta(q^0+\omega^l_k-\omega^n_p)} \right]\label{eq.Impi.B.d3k}
	\end{eqnarray}
where $\omega^l_k=\sqrt{k_z^2+m_l^2}, ~\omega^n_p=\sqrt{p_z^2+m_n^2},~ f_k^l=f_\text{BE}(\omega_k^l),~ f_p^n=f_\text{BE}(\omega_p^n)$ and  $\Ntil^{\mu\nu}_{nl}(q,k_\parallel,k_\perp)=4(-1)^{n+l}e^{-\alpha_k-\alpha_p}L_l(2\alpha_k)L_n(2\alpha_p)N^{\mu\nu}$. The expression for $\text{Re}~\Pibar^{\mu\nu}_{\rm~Vac}(q,eB)$ is given in Appendix~\ref{Apendx.RePi.VccB}. Now performing the $d^2k_\perp$ integrations of Eqs.~\eqref{eq.repi.B.d3k} and \eqref{eq.Impi.B.d3k}, one obtains
\begin{eqnarray}
	\text{Re}~\Pibar^{\mu\nu}(q;T,eB) &=&\text{Re}~\Pibar^{\mu\nu}_{\rm~Vac}(q,eB) + \sum_{n=0}^{\infty}~\sum_{l=0}^{\infty}~\int_{-\infty}^{\infty}\frac{dk_z}{2\pi}~\mathcal{P}\bigg[
	\frac{f_k^l}{2\omega^l_k} \bigg\{\frac{N^{\mu\nu}_{nl}(q,k^0=-\omega^l_k,k_z)}{(q^0-\omega^l_k)^2-(\omega^n_p)^2}
	+\frac{N^{\mu\nu}_{nl}(q,k^0=\omega^l_k,k_z)}{(q^0+\omega^l_k)^2-(\omega^n_p)^2} \bigg\} \nn \\ 
	&& +\frac{f^n_p}{2\omega^n_p} \bigg\{\frac{N^{\mu\nu}_{nl}(q,k^0=-q^0-\omega^n_p,k_z)}{(q^0+\omega^n_p)^2-(\omega^l_k)^2} 
	+ \frac{N^{\mu\nu}_{nl}(q,k^0=-q^0+\omega^n_p,k_z)}{(q^0-\omega^n_p)^2-(\omega^l_k)^2} \bigg\} \bigg], \label{eq.repi.B} \\
	\text{Im}~\Pibar^{\mu\nu}(q;T,eB)&=&-\tanh\FB{\frac{|q^0|}{2T}}\pi\sum_{n=0}^{\infty}~\sum_{l=0}^{\infty}\int_{-\infty}^{+\infty}\frac{dk_z}{2\pi}\frac{1}{4\omega^l_k\omega^n_p} \nn \\
	&&\hspace{-1cm} \TB{  (1+f^l_k+f^n_p+2f^l_kf^n_p) \SB{N^{\mu\nu}_{nl}(q,k^0=-\omega^l_k,k_z)\delta(q^0-\omega^l_k-\omega^n_p) + N^{\mu\nu}_{nl}(q,k^0=\omega^l_k,k_z)\delta(q^0+\omega^l_k+\omega^n_p)} \right. \nn \\
	&&\hspace{-1cm} \left.+(f^l_k+f^n_p+2f^l_kf^n_p)\SB{N^{\mu\nu}_{nl}(q,k^0=-\omega^l_k,k_z)\delta(q^0-\omega^l_k+\omega^n_p) + N^{\mu\nu}_{nl}(q,k^0=\omega^l_k,k_z)\delta(q^0+\omega^l_k-\omega^n_p)} } \label{ImPi1}
\end{eqnarray}
where
\begin{equation}\label{Eq.Numununl}
	N^{\mu\nu}_{nl}(q,k_\parallel)=\int\frac{d^2k_\perp}{\FB{2\pi}^2}\Ntil^{\mu\nu}_{nl}(q,k_\parallel,k_\perp).
\end{equation}
The $dk_z$ integration in Eq.~\eqref{ImPi1} can now be performed using the Dirac delta function and we get, 
	\begin{eqnarray}
	\text{Im}~\Pibar^{\mu\nu}(q;T,eB)&=&-\tanh\FB{\frac{|q^0|}{2T}}\sum_{n=0}^{\infty}~\sum_{l=0}^{\infty}\frac{1}{4\lambda^{1/2}(q_\parallel^2,m_l^2,m_n^2)}\sum_{k_z \in \{k_z^\pm\}} 
	\Big[  (1+f^l_k+f^n_p+2f^l_kf^n_p) \nn \\
	&& \times \Big\{ N^{\mu\nu}_{nl}(q,k^0=-\omega^l_k,k_z) \Theta \Big(q^0-\sqrt{q_z^2+\FB{m_l+m_n}^2} \Big)
	+ N^{\mu\nu}_{nl}(q,k^0=\omega^l_k,k_z)\Theta \Big(-q^0-\sqrt{q_z^2+\FB{m_l+m_n}^2}\Big) \Big\}\nn\\
	&& +(f^l_k+f^n_p+2f^l_kf^n_p) \Big\{N^{\mu\nu}_{nl}(q,k^0=-\omega^l_k,k_z)\Theta\FB{q^0-\text{min}(q_z, E^{\pm})} \Theta \Big( -q^0+\text{max}(q_z, E^{\pm})\Big)\nn\\&& + N^{\mu\nu}_{nl}(q,k^0=\omega^l_k,k_z)\Theta\FB{-q^0-\text{min}(q_z, E^{\pm})} \Theta \Big( q^0+\text{max}(q_z, E^{\pm})\Big) \Big\}
	\Big] \label{ImPi3}
	\end{eqnarray}
	where $\lambda(x,y,z)=x^2+y^2+z^2-2xy-2yz-2zx$ is the K\"all\'en function, 
	$k_z^{\pm}=\frac{1}{2q_\parallel^2}\TB{-yq_z\pm|q^0|\lambda^{1/2}(q_\parallel^2, m_l^2, m_n^2)}$, ~$y=\FB{q_{\parallel}^2+m_l^2-m_n^2}$, and $E^{\pm}=\frac{m_l-m_n}{|m_l\pm m_n|}\sqrt{q_z^2+\FB{m_l\pm m_n}^2}$.

Eq.~\eqref{ImPi1} contains four Dirac delta functions similar to Eq.~\eqref{ImPi2} representing the Unitary and Landau cuts. Because of dimensional reduction, they contain only the longitudinal dynamics. Unlike the vanishing $eB$ case, here a non-trivial Landau cut contribution may appear in the (physical) time-like kinematic domain (even if the loop-particles have the same masses). This happens when the pions in the loop occupy different Landau levels. Physically this means that a rho-meson can be absorbed by means of scattering with a pion in lower Landau level producing a pion in higher Landau level in the final state (and the time reversed process). The contributions from Unitary-I and Unitary-II are non-vanishing in the kinematic regions $\sqrt{q_z^2+4(m_\pi^2+eB)}<q^0<\infty$ and -$\infty<q^0<-\sqrt{q_z^2+4(m_\pi^2+eB)}$ respectively. On the other hand, the non-vanishing kinematic domain for both the Landau cuts is $|q^0|<\text{max}(q_z, E^{\pm})$.
\begin{figure}[h]
	\includegraphics[scale=0.9]{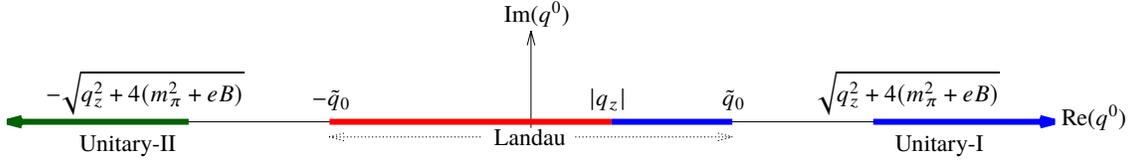}
	\caption{Analytic structure of $\Pibar(q^0,q_z;T,eB)$ in complex plane of $q^0$ for a given value of $q_z$. Here, $\widetilde{q}_0=\sqrt{q_z^2+(\sqrt{m_\pi^2+eB}-\sqrt{m_\pi^2+3eB}~)^2}$. Unitary-I and some portion of the Landau cuts (denoted by blue line) corresponds to the domain of physical dilepton production.}
	\label{KDM2}
\end{figure}

The analytic structure of the self-energy is easier to understand if we consider the case of $q_\perp=0$. In this situation, Eq.~\eqref{ImPi3} will be simplified as $l$ will lie between $(n-1)$ to $(n+1)$ for a given value of $n$, so that
\begin{eqnarray}\label{Nmununl}
N^{\mu\nu}_{nl}(q_\parallel,q_\perp=0,k_\parallel)&=&4g^2_{\rho\pi\pi}(-1)^{n+1}\frac{eB}{8\pi} \Big[ 
\SB{k^\mu_{\parallel}k^\nu_{\parallel}q^4_{\parallel}+q^\mu_{\parallel}q^\nu_{\parallel}\FB{q_{\parallel}.k_{\parallel}}^2
	-\FB{k^\mu_{\parallel}q^\nu_{\parallel}+k^\nu_{\parallel}q^\mu_{\parallel}}q^2_{\parallel}\FB{q_{\parallel}.k_{\parallel}}}\delta^n_l \nn \\ 
&&\hspace{3cm} + q_\parallel^4g^{\mu\nu}_\perp\frac{eB}{4}\SB{n\delta^{n-1}_l-\FB{2n+1}\delta^n_l+\FB{n+1}\delta^{n+1}_l} \Big]
\end{eqnarray}	
As a result, kinematic domain of Landau cuts will be modified and the non-vanishing region for both the Landau cuts is $|q^0|<\sqrt{q_z^2+(\sqrt{m_\pi^2+eB}-\sqrt{m_\pi^2+3eB}~)^2}$. The cut structure of the thermo-magnetic self-energy function is shown in Fig.~\ref{KDM2}.

In order to solve the Dyson-Schwinger equation for the complete $\rho^0$-propagator, we use the following Lorentz decomposition of the self-energy 
at finite temperature under external magnetic field~\cite{Ghosh:2019fet}
\begin{equation}
	\Pibar^{\mu\nu}(T,eB)=\Pi_A P_A^{\mu\nu}+\Pi_B P_B^{\mu\nu}+\Pi_C P_C^{\mu\nu}+\Pi_L P_L^{\mu\nu} \label{eq.PiB.dec}
\end{equation}
where the basis tensors are
\begin{eqnarray}
	P_A^{\mu\nu}&=&\FB{g^{\mu\nu}-\frac{q^\mu q^{\nu}}{q^2}-\frac{\util^\mu\util^\nu}{\util^2}-\frac{\btil^\mu\btil^\nu}{\btil^2}}, \label{eq.P1}\\
	P_B^{\mu\nu}&=&\frac{\btil^\mu\btil^\nu}{\btil^2},~~ P_L^{\mu\nu}~=~\frac{\util^\mu\util^\nu}{\util^2},\\
	P_C^{\mu\nu}&=&\frac{1}{\sqrt{\util^2\btil^2}}\FB{\util^\mu\btil^\nu+\util^\nu\btil^\mu}. \label{eq.Q}
\end{eqnarray}
In Eqs.~\eqref{eq.P1}-\eqref{eq.Q}, $\btil^\mu = b^\mu-\FB{\frac{q\cdot b}{q^2}}q^\mu-\FB{\frac{b\cdot\util}{\util^2}}\util^\mu$ 
where, $b^\mu = \frac{1}{2B}\varepsilon^{\mu\nu\alpha\beta} F^\text{ext}_{\nu\alpha}u_\beta$ in which $F^\text{ext}_{\nu\alpha}$ is the electromagnetic field strength tensor corresponding 
to the external magnetic field. It may be noted that in LRF, $b^\mu_\text{LRF}\equiv(0,\hat{\bm{z}})$ points along the direction of external magnetic field. Also, the vector $\btil^\mu$ is orthogonal 
to both $q^\mu$ and $\util^\mu$. The form factors in Eq.~\eqref{eq.PiB.dec} comes out to be
\begin{eqnarray}
	\Pi_L&=&\frac{1}{\util^2}u_\mu u_\nu\Pibar^{\mu\nu}\\
	\Pi_C&=&\frac{1}{\sqrt{\util^2\btil^2}} \Big\{u_\mu b_\nu\Pibar^{\mu\nu}-(b\cdot\util)\Pi_L\Big\} \\
	\Pi_B&=&\frac{1}{\btil^2} \Big\{ b_\mu b_\nu \Pibar^{\mu\nu}+\frac{\FB{b\cdot\util}^2}{\util^2}\Pi_L-2\frac{b\cdot\util}{\util^2}u_\mu b_\nu \Pibar^{\mu\nu} \Big\}\\
	\Pi_A&=&\FB{g_{\mu\nu}\Pibar^{\mu\nu}-\Pi_L-\Pi_B}\label{MFF}.
\end{eqnarray}
Thus, using Eqs.~\eqref{rhoprop} and \eqref{eq.PiB.dec}, Eq.~(\ref{Dyson}) is solved to get the complete thermo-magnetic $\rho^0$-meson propagator as
\begin{eqnarray} \label{Eq_D_B}
	\overline{D}^{\mu\nu}(T,eB)&=&\frac{P_A^{\mu\nu}}{q^2-m_\rho^2+\Pi_A} 
	+\frac{(q^2-m_\rho^2+\Pi_L)P_B^{\mu\nu}}{\FB{q^2-m^2_\rho+\Pi_B}\FB{q^2-m_\rho^2+\Pi_L}-\Pi^2_C}  
	-\frac{\Pi_C P_C^{\mu\nu}}{\FB{q^2-m_\rho^2+\Pi_L}\FB{q^2-m_\rho^2+\Pi_B}-\Pi_C^2}  \nn \\
	&& +\frac{(q^2-m_\rho^2+\Pi_B)P_L^{\mu\nu}}{\FB{q^2-m_\rho^2+\Pi_B}\FB{q^2-m_\rho^2+\Pi_L}-\Pi^2_C}  - \frac{q^\mu q^\nu}{q^2m_\rho^2}.  \label{eq.Dbar.B}
\end{eqnarray}

It turns out that the consideration of vanishing transverse momentum $q_\perp=0$ simplifies the form factors significantly; in particular, 
we get $\Pi_A(q_\perp=0)=\Pi_B(q_\perp=0)=\Pi_T$ (say) and $\Pi_C(q_\perp=0)=0$. The imaginary part of $\overline{D}^{\mu\nu}$ in Eq.~\eqref{eq.Dbar.B} gives 
the thermo-magnetic spectral function of rho meson as $\mathcal{A}(q;T,eB)=-\frac{1}{3}g^{\mu\nu}\text{Im}~\overline{D}_{\mu\nu}$ which is to be plugged into Eq.~\eqref{DPR2} 
to calculate the DPR. We emphasize that in case of non-zero magnetic field, physical dileptons (having $q^0>0$ and $q^2>0$) can be produced from both the Unitary-I and Landau cuts 
(as shown by blue region in Fig.~\ref{KDM2}) where the spectral function is non-zero.

%~~~~~~~~~~~~~~~~~~~~~~~~~~~~~~~~~~~~~~~~~~~~~~~~~~~~~~~~~~~~~~~~~~~~~~~~~~~~~~~~~~~~~~~~~~~~~~~~~~~~~~~~~~~~~~~~~~~~~~~~~~~
\section{Numerical Results}\label{Numerical}
In this section, numerical results for several quantities, such as, imaginary part of $ \rho^0 $ self-energy and complete $ \rho^0 $-propagator, dilepton production rate \textit{etc} are presented in different physical scenarios. It should be noted that while calculating the components of $\IM~\overline{\Pi}^{\munu}$ in presence of non-zero magnetic field, we have to perform sum over infinite number of Landau levels (see Eq.~\eqref{ImPi3}). However, for all numerical results, we have taken up to 500 Landau levels which ensures the convergence of the sum. We present our results for $T=130$ and $160$ MeV which are representative temperatures of the hadronic phase. Since the hadronic phase is formed in the late stage of the evolution, a weaker magnetic field $eB=0.02~\rm GeV^2$ has been considered in the numerical results. However, some higher values e.g. $eB=0.03,~0.05~\rm GeV^2$ are also considered to show the dependency of the magnetic field on the numerical results as our calculation is valid for arbitrary values of magnetic field. The representative values  $q_\perp=150$ MeV and $q_z = 150$ MeV are chosen which are of the same order as the temperature. We also show results of DPR for different values of $q_\perp$  and $ q_z $. We have taken rest mass of pion as $m_\pi=140~\rm MeV$.
 
 \begin{figure}[h] 
 	\includegraphics[angle = -90, scale=.34]{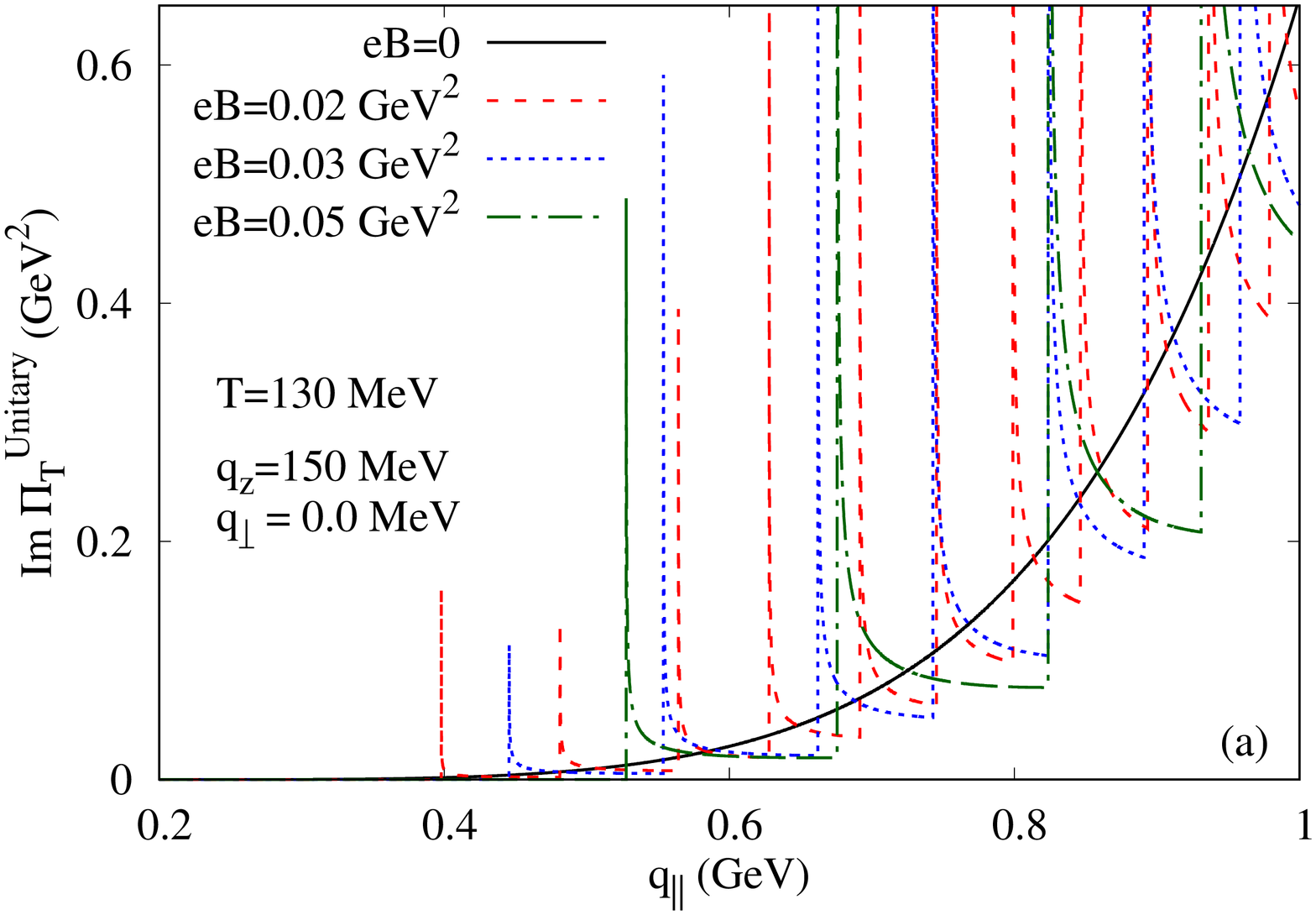}
 	\includegraphics[angle = -90, scale=.34]{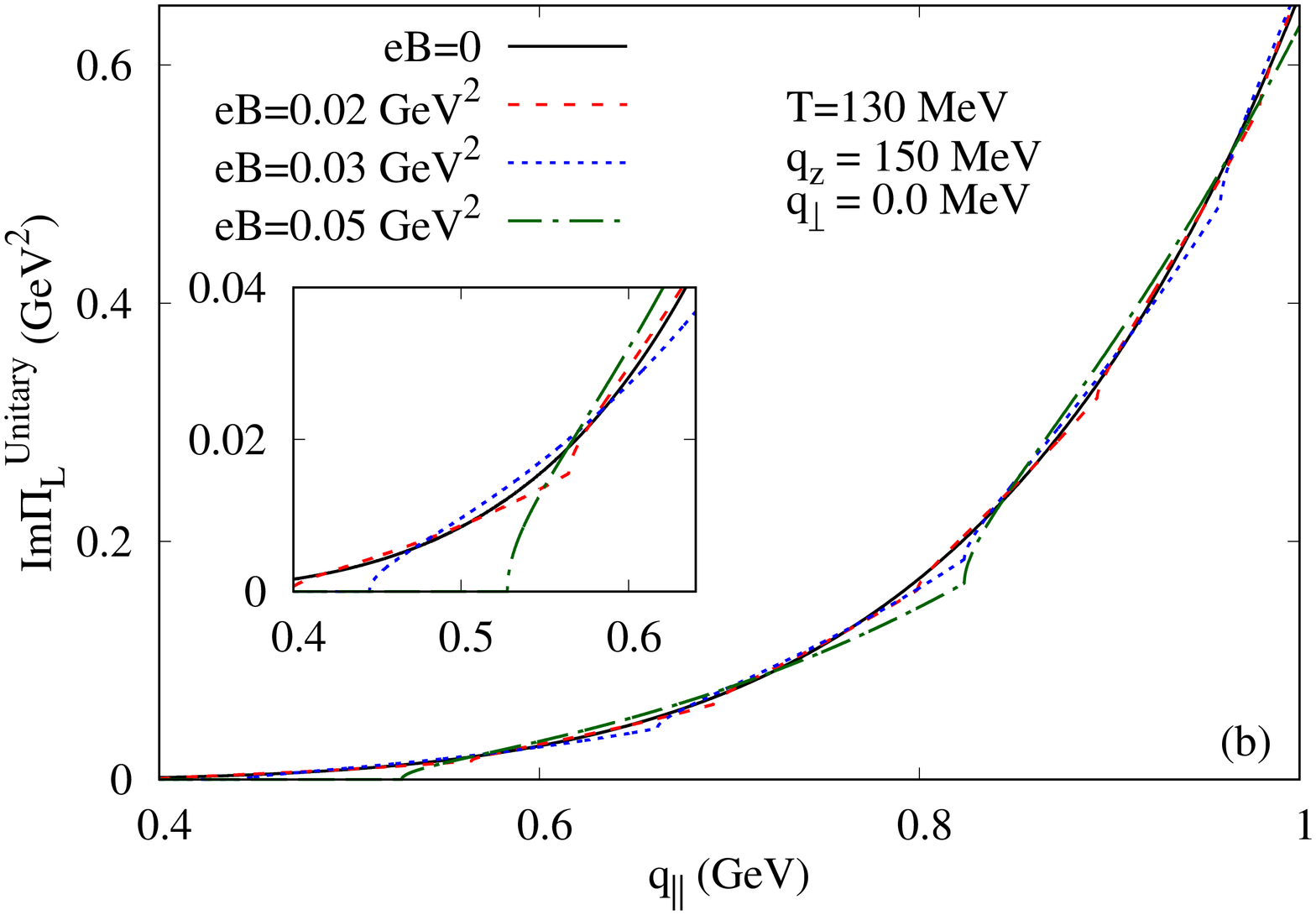}
 	\includegraphics[angle = -90, scale=.34]{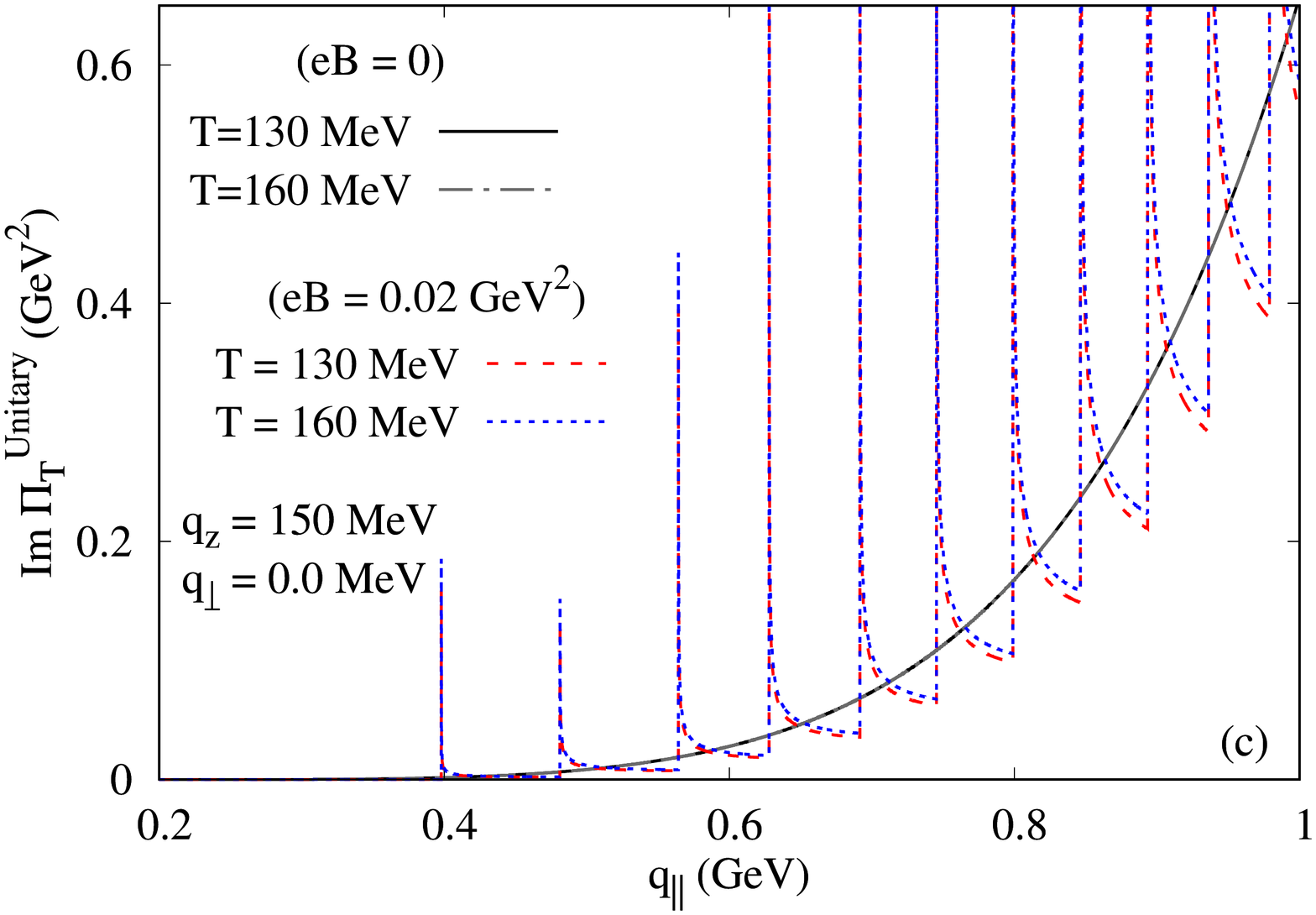}
 	\includegraphics[angle = -90, scale=.34]{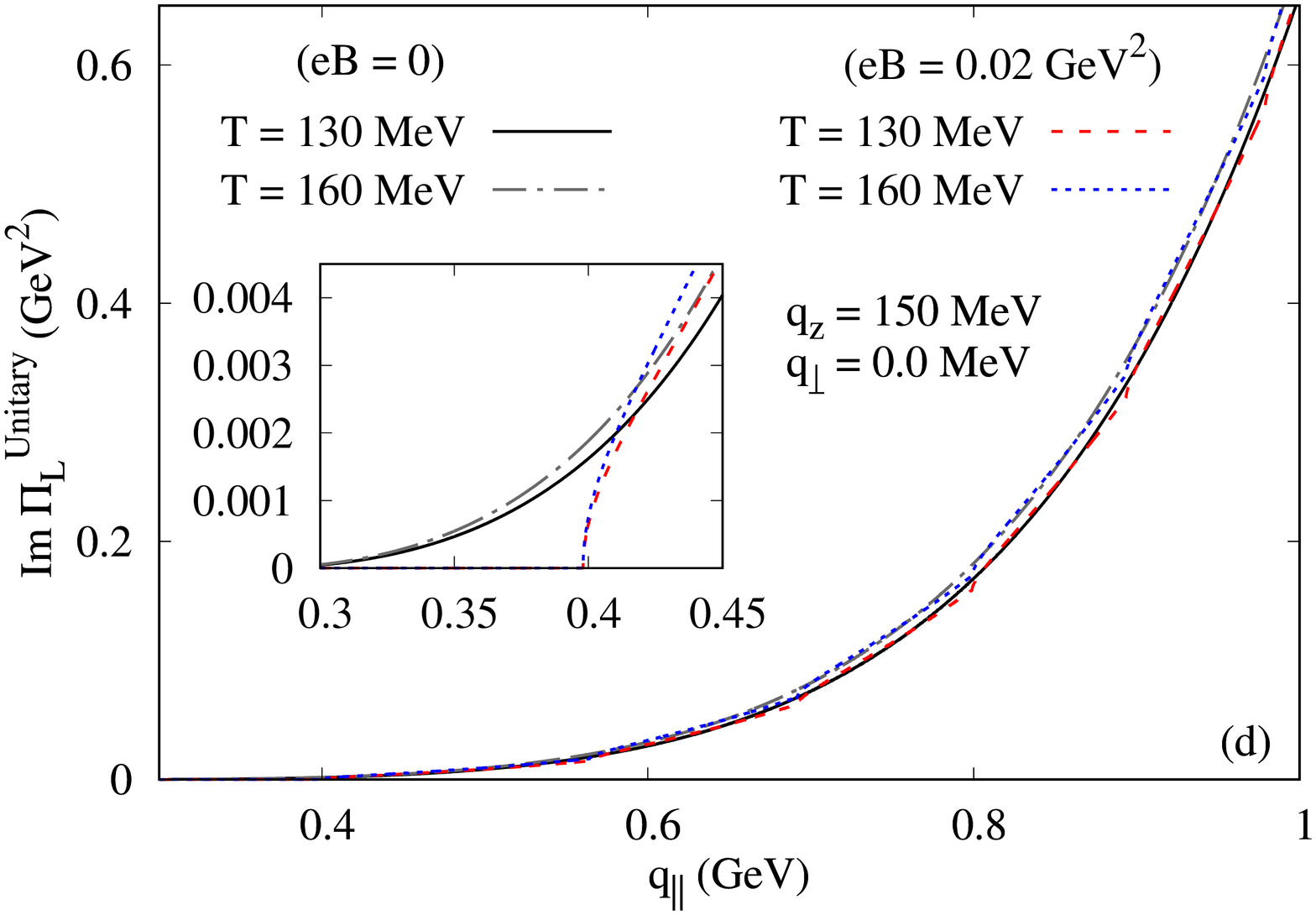}		
 	\caption{Unitary cut contributions in (a) $ \IMA $ and (b) $ \IMB $ at $T=130$ MeV, (c) $ \IMA $ and (d) $ \IMB $ at $T=130$ and $160$ MeV as a function of $ q_\parallel $ for $q_z=150$ MeV, $q_\perp=0.0$ for different $eB$ values.}
 	\label{Fig_ImPi}
 \end{figure}
\begin{figure}[h] 
	\includegraphics[angle = -90, scale=.34]{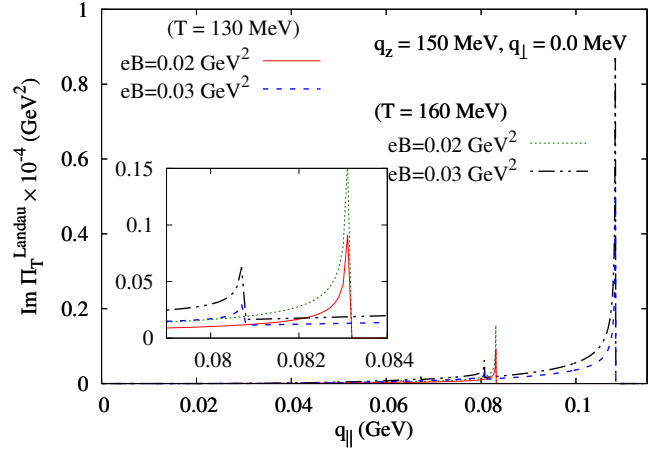}		
	\caption{Landau cut contributions in $ \IMA $ as a function of ${q_\parallel}$ at $q_z=150$ MeV for $T=130$ and $160$ MeV for different $ eB $ values.}
	\label{Fig_ImPiL}
\end{figure}
\begin{figure}[h] 
	\includegraphics[angle = -90, scale=.35]{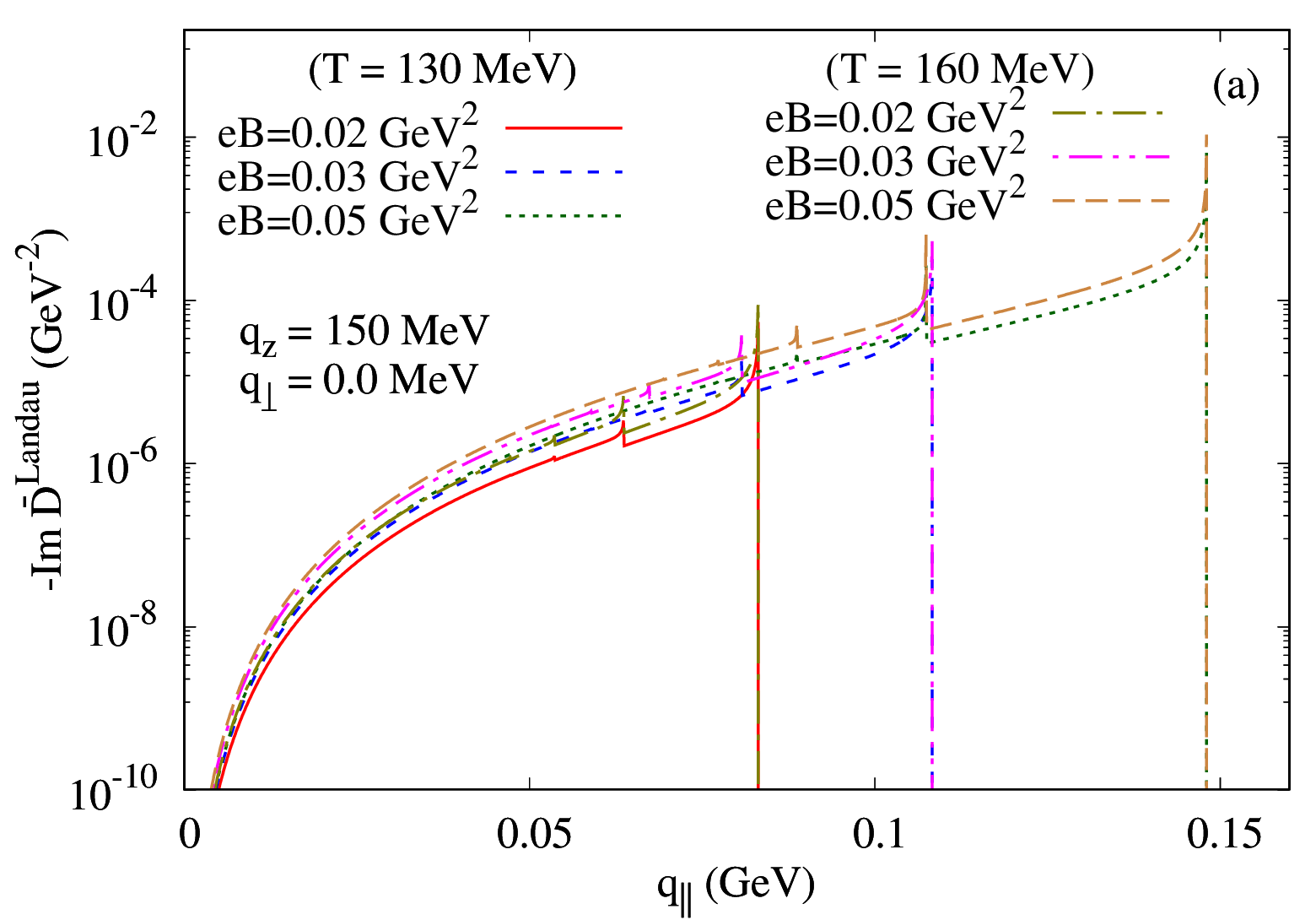}
	\includegraphics[angle = -90, scale=.35]{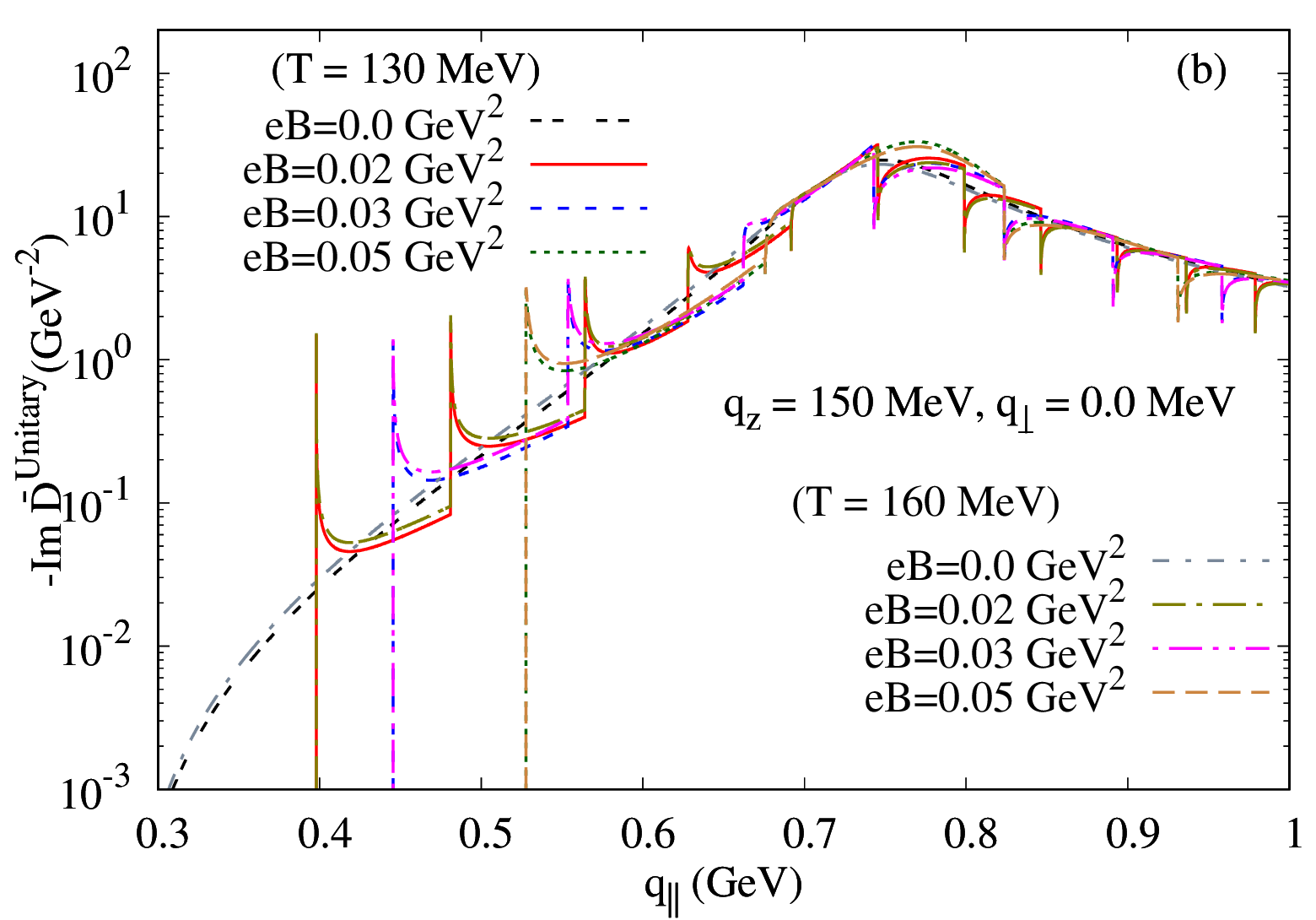}
	\caption{The variation of (a) Landau cut and (b) Unitary cut contributions in the complete $\rho^0$ propagator as a function of ${q_\parallel}$  at $q_z=150$ MeV for different $ eB $-values at $T=130$ and $160$ MeV.}
	\label{Fig_D}
\end{figure}

We first consider the case of zero transverse momentum, i.e, $q_\perp=0$ and longitudinal momentum $q_z=150$ MeV in Figs.~\ref{Fig_ImPi},~\ref{Fig_ImPiL} and \ref{Fig_D}. In Figs.~\ref{Fig_ImPi}(a) and (b), we have shown the contribution of the Unitary cut in  $ \IMA $ and $ \IMB  $ respectively as function of $ \qL $ at temperature $ T = 130  $~MeV for different values of external magnetic field. From Fig.~\ref{Fig_ImPi}(a), it is evident that $ \IMAU $  consists of spike-like structures separated from each other by a finite value for non-zero values of $ eB $ and the form factor oscillates about the $ eB = 0 $ plot. The appearance of these spikes is due to the so called “threshold singularities” at each Landau level~\cite{Ghosh:2017rjo,Chakraborty:2017vvg,Ghosh:2018xhh,Ghosh:2019fet}. Mathematically this can be understood from Eq.~\eqref{ImPi3} where the K\"all\'en function appearing in the denominator goes to zero at each threshold of the Unitary cut defined in terms of the unit step functions. As discussed below Eq.~\eqref{ImPi3}, the threshold for the Unitary cut for different values of $ eB $ can be determined from the following condition
\begin{equation}\label{Eq_UCut}
\qL > 2 \sqrt{m_\pi^2 + eB}  .
\end{equation}
 Furthermore, Eq.~\eqref{Eq_UCut} also predicts that the threshold of the Unitary cut should shift towards higher values of $ \qL $ as $ eB $ increases which is evident from 
 Fig.~\ref{Fig_ImPi}(a). In Fig~\ref{Fig_ImPi}(b), we have plotted $ \IMBU $ as function of $ \qL $. Unlike $ \IMAU $, $ \IMBU $ does not contain any spike-like structure. 
 This is due to an extra factor of K\"all\'en function coming from the component ${N}_{nl}^{00}$ (which contributes to $ \IMB $), canceling the same in the denominator of Eq.~\eqref{ImPi3}.  
 Similar to  $ \IMAU $, it can also be seen that with non-zero values of $eB$, $ \IMBU $ is approximately same as the $ eB= 0 $ curve but the oscillation frequency is much smaller as compared to $ \IMAU $. Moreover, analogous to Fig.~\ref{Fig_ImPi}(a), the threshold of the unitary cut moves towards higher values of $ \qL $ with the increase in magnetic field as clearly shown in the inset plot. In Figs.~\ref{Fig_ImPi}(c) and (d) we have presented the variation of $ \IMAU $ and $ \IMBU $ with $ \qL $ for a fixed value of the background field ($ eB = 0.02$ GeV$^2$) for two different values of $ T $. In both the figures, the corresponding curves for $ eB = 0 $ case are shown for comparison. It is evident that for different values of temperature the qualitative behavior of both $ \IMAU $ and $ \IMBU $ remains similar. However, there is an increase in the magnitude of both the contribution of the Unitary cut in  $ \IMA $ and $ \IMB  $ for higher value of $ T $ owing to the availability of larger thermal phase space.

%~~~~~~~~~~~~~~~~~~~~~~~~~~~~~~~~~~~~~~~~~~~~~~~~~~~~~~~~~~~~~~~~~~~~~~~~~~~~~~

As pointed out earlier while discussing the detailed analytic structure of the self-energy of $ \rho^0 $-meson, a nontrivial Landau cut contribution might be generated in the presence of external magnetic field even if the loop particles carry equal mass. In this case, the nonzero Landau cut contribution will only appear in $ \IMA $ as can be understood from Eq.~\eqref{Eq_Nmumu_B} or Eq.~\eqref{Nmununl} where $g_{\mu\nu}{N}^{\mu\nu}_{nl}$ (which contributes to $ \IMA $) contains two additional Kronecker delta functions $\delta^{n\mp l}_l$ as well as $\delta^n_l$. However, from Eq.~\eqref{Eq_N00_B} or Eq.~\eqref{Nmununl}, it is evident that such feature is absent in the expression of ${N}_{nl}^{00}$ (which contributes in $ \IMB $). In Fig.~\ref{Fig_ImPiL}, we have depicted the contribution of the Landau cuts in $ \IMA $ as a function $ \qL $ at $ T = 130 $ and $ 160 $~MeV for different values of magnetic field. Comparing with Fig.~\ref{Fig_ImPi} (a) and (c), it can be observed that the magnitude of $ \IMAL $ is $\sim\times 10^{-4} $ smaller compared to $ \IMAU $ and the Landau cut contributions also contain the threshold singularities.  Now again as discussed below Eq.~\eqref{Nmununl}, the kinematic domain for the Landau cut can be determined from the following condition
\begin{equation}\label{Eq_LCut}
\qL < {\sqrt{m_\pi^2+3eB}-\sqrt{m_\pi^2+eB}} .
\end{equation}
This explains the fact that the Landau cut contributions in $ \IMA $ extend towards higher values of $ \qL $ with the increase in $ eB $  as evident from Fig.~\ref{Fig_ImPiL}. Moreover, for higher $ T $ value, due to the enhancement of the thermal factor, which in turn increases the available thermal phase space, the magnitude of $ \IMAL $ is larger. It should be noted that both the Unitary and Landau cut threshold, determined by Eqs.~\eqref{Eq_UCut} and \eqref{Eq_LCut} respectively, are independent of the the temperature of the medium. All the observations made in Figs.~\ref{Fig_ImPi} and \ref{Fig_ImPiL} are in agreement with the results obtained in~\cite{Ghosh:2019fet}.
\begin{figure}[h]
	\includegraphics[angle = -90, scale=.35]{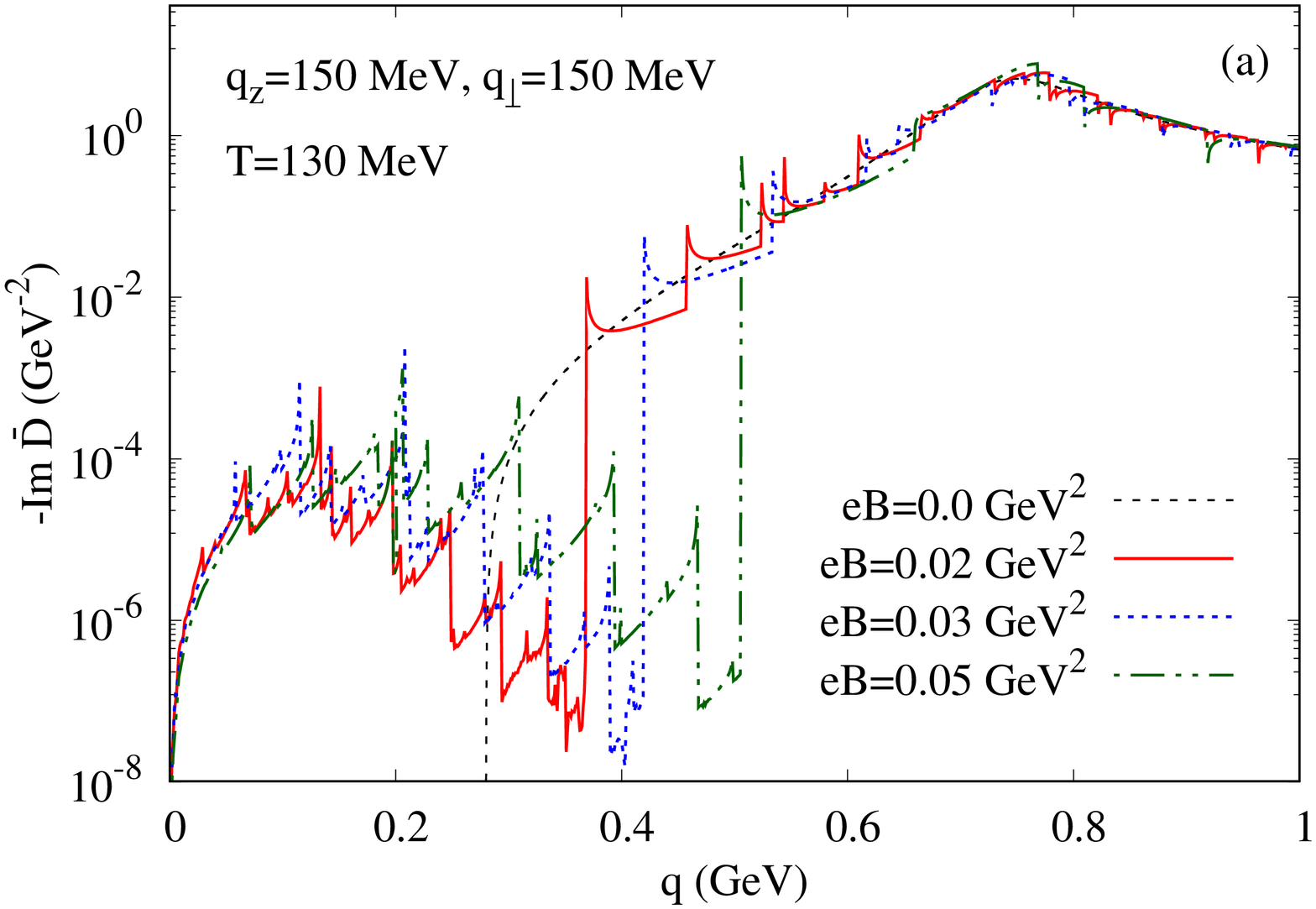}
	\includegraphics[angle = -90, scale=.35]{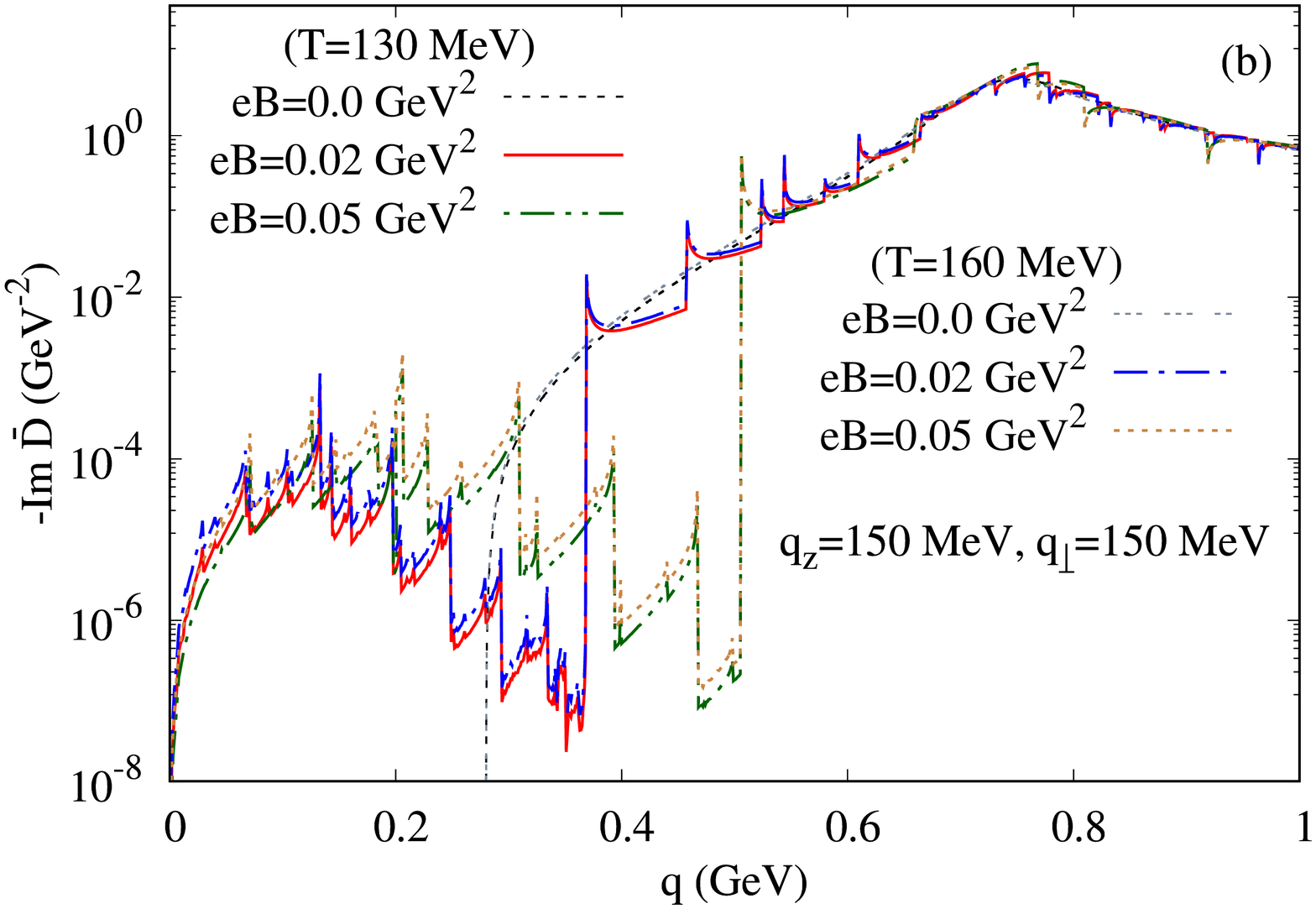}
	\caption{The variation of the complete $\rho^0$ propagator as a function of $q$  at $q_z=150$ MeV, $q_\perp=150$ MeV for different $ eB $-values at (a) $T=130$ MeV and (b) $160$ MeV.}
	\label{Fig.ImD}
\end{figure}

It is clear from Eq.~\eqref{Eq_D_B} and discussion below that, for the vanishing transverse momentum of $ \rho^0 $, the complete propagator of $ \rho^0 $-meson consists of three structure factors in a magnetized medium. Out of these, two are found to be degenerate, implying two distinct structure factors for the propagation of $ \rho^0 $. A detailed study of these structure factors can be found in~\cite{Ghosh:2019fet}. In Figs.~\ref{Fig_D}(a) and (b), we illustrate the variation of Landau and Unitary cut contributions respectively in the complete $\rho^0$-meson propagator as a function of $\qL$ for different $ eB $-values at $T=130$ and 160 MeV, $q_\perp=0.0$ and $q_z = 150 $ MeV. Both the plots contain spike-like structure owing to the threshold singularities at each Landau level as discussed earlier. The increase (decrease) in Landau (Unitary) cut threshold with increase in magnetic field can be explained in a similar fashion using Eq.~\eqref{Eq_LCut} (Eq.~\eqref{Eq_UCut}). From Fig~\ref{Fig_D}(b), it can be seen that for a particular value of background field, the overall width of {$ \IM ~\overline{D}^{\rm Unitary} $} broadens with the increase in temperature. This corresponds to the enhancement of the decay process $\rho^0$$\rightarrow$$\pi^+\pi^-$ in that medium indicating that  $\rho^0$-meson becomes  more unstable at high temperature (see Ref.~\cite{Ghosh:2019fet} for more details).
For non-zero transverse momentum of $\rho^0$ meson, the complete $\rho^0$ propagator has four structure factors in thermo-magnetic medium (see Eq.~\eqref{Eq_D_B}). Fig.~\ref{Fig.ImD}(a) depicts $ \IM ~\overline{D} $ as a function of $\sqrt{q^2}$ at non-zero value of $q_\perp$ and $q_z$ for different values of background magnetic field.  With finite value of $q_\perp$, the threshold of Unitary cut shifts towards the lower invariant mass and the threshold of Landau cut shifts towards the higher invariant mass region leading to a continuous spectrum in $ \IM ~\overline{D}$ which is an interesting observation at non-vanishing transverse and longitudinal momenta of $\rho^0$. 
There is an overall increase in $ \IM ~\overline{D}$ with increasing temperature in lower $\sqrt{q^2}\FB{<~\sqrt{4(m_\pi^2+eB)+q_\perp^2}}$ region. However, the nature of $ \IM ~\overline{D}$ (in Fig.~\ref{Fig.ImD}(b)) is same as $ \IM ~\overline{D}^{\rm Unitary}$ (in Fig.~\ref{Fig_D}(b)) in higher $\sqrt{q^2}\FB{>\sqrt{4(m_\pi^2+eB)+q_\perp^2}}$ domain.

\begin{figure}[h]
	\includegraphics[angle = -90, scale=.35]{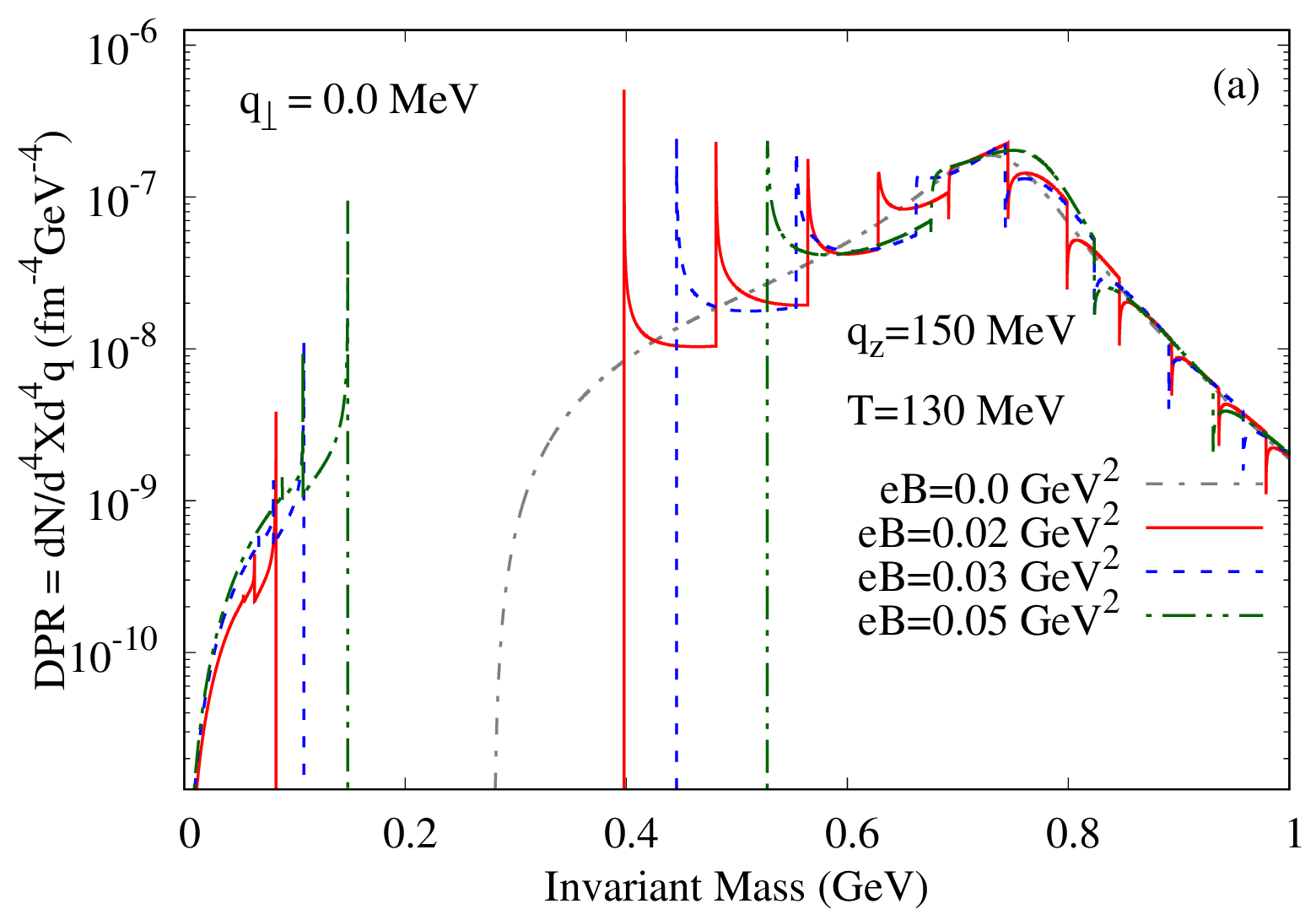}
	\includegraphics[angle = -90, scale=.35]{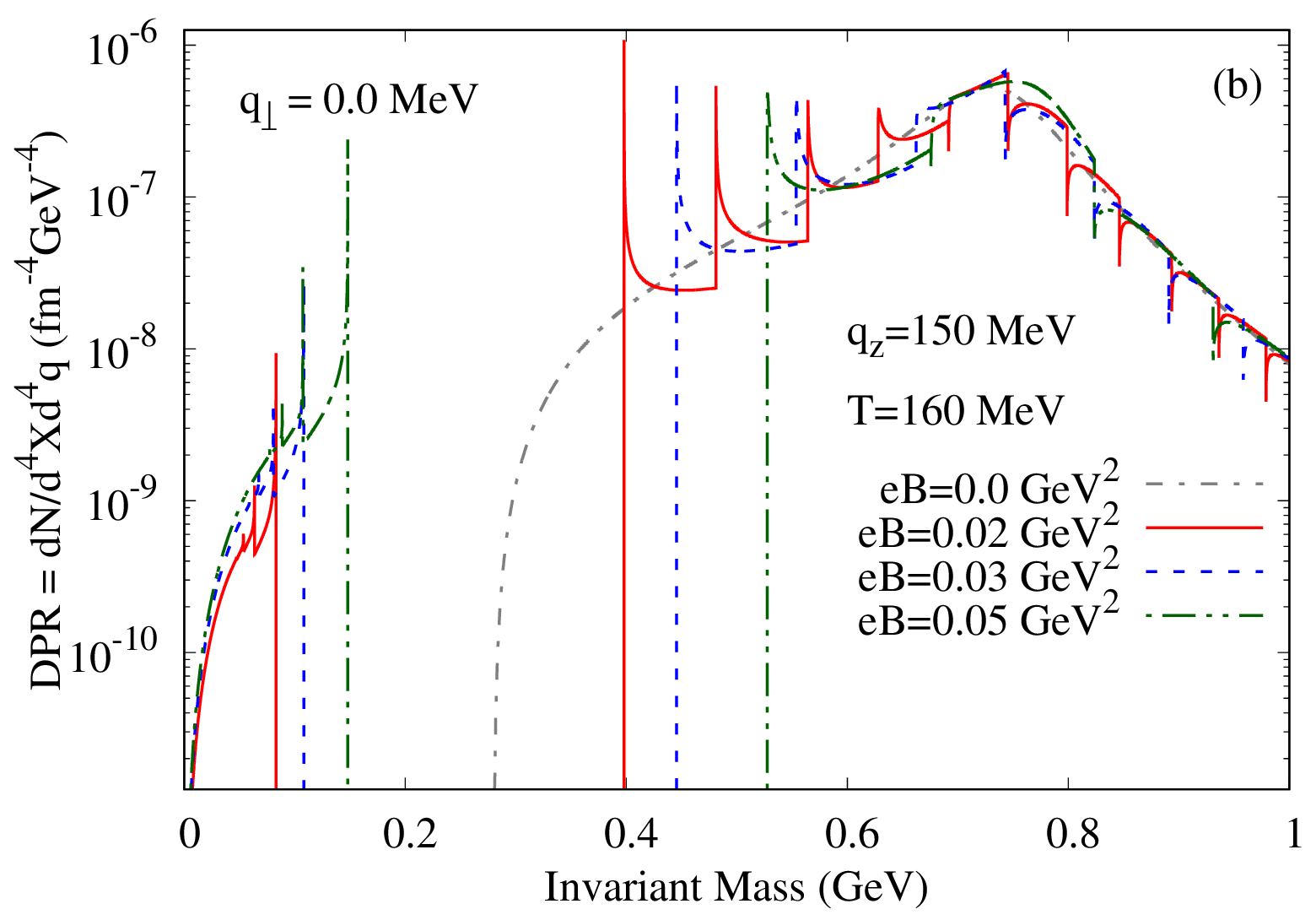}
	\includegraphics[angle = -90, scale=.35]{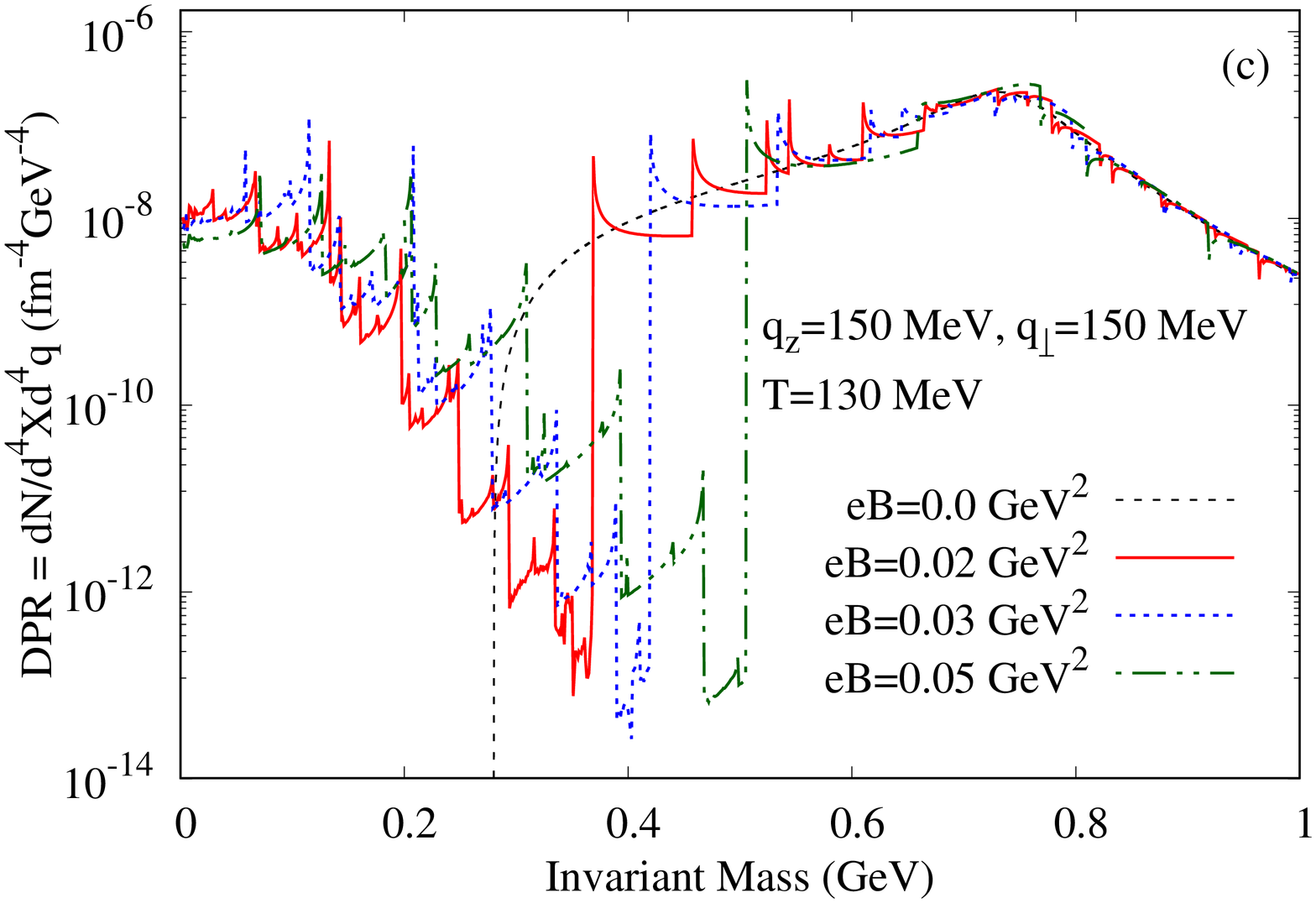}
	\includegraphics[angle = -90, scale=.35]{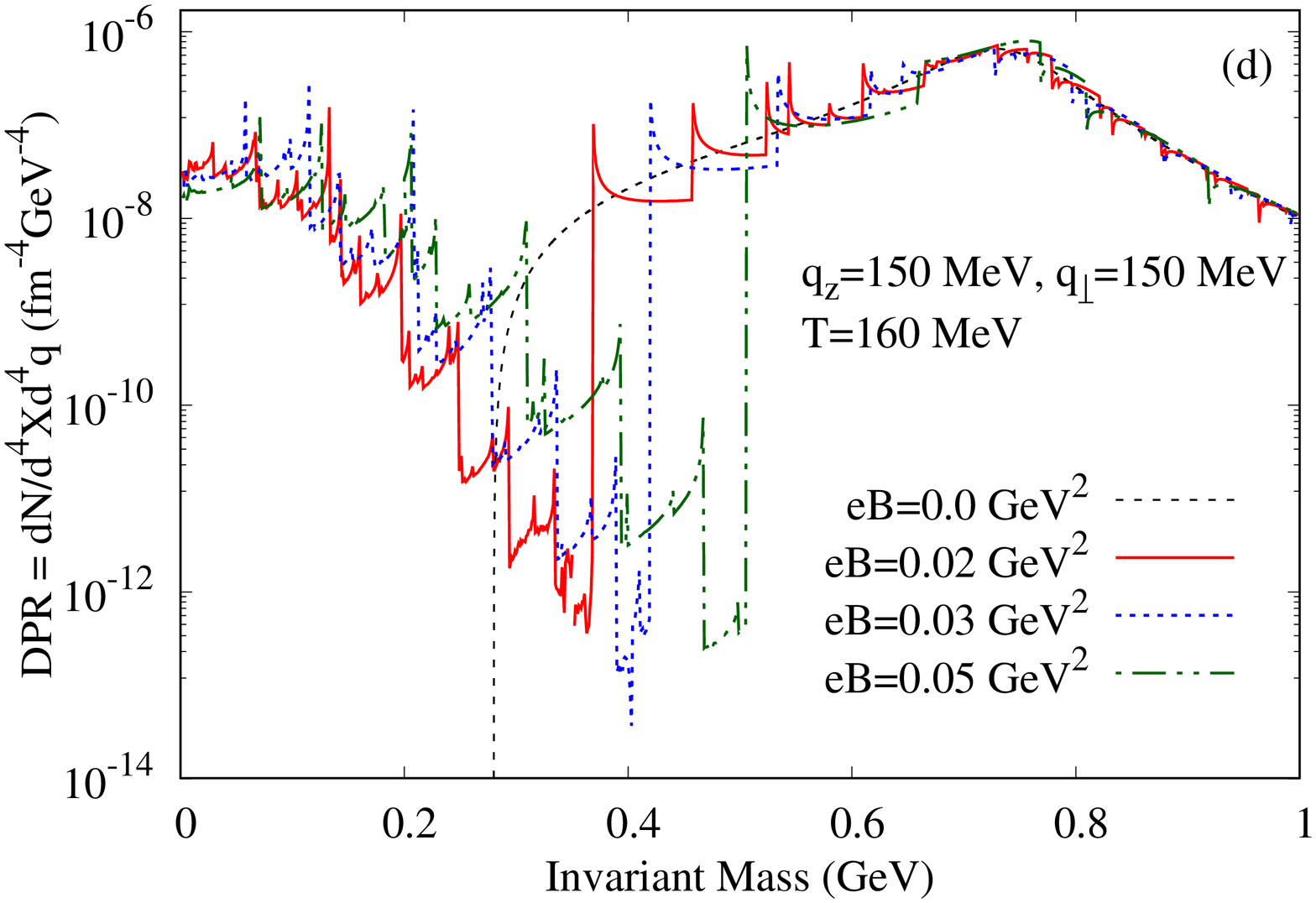}
	\caption{Dilepton production rate (DPR) as a function of the invariant mass at $q_z=150$ MeV for different values background magnetic field for  $ q_\perp = 0  $ at (a) $T=130$ and (b) $T=160$ MeV, for  $ q_\perp = 150  $ MeV at (c) $T=130$  and (d) $T=160$ MeV (corresponding $ eB = 0 $ curves (grey dotted line) are also shown for comparison)}
	\label{Fig_DPR}
\end{figure}

\begin{figure}[h]
	\includegraphics[angle = -90, scale=.35]{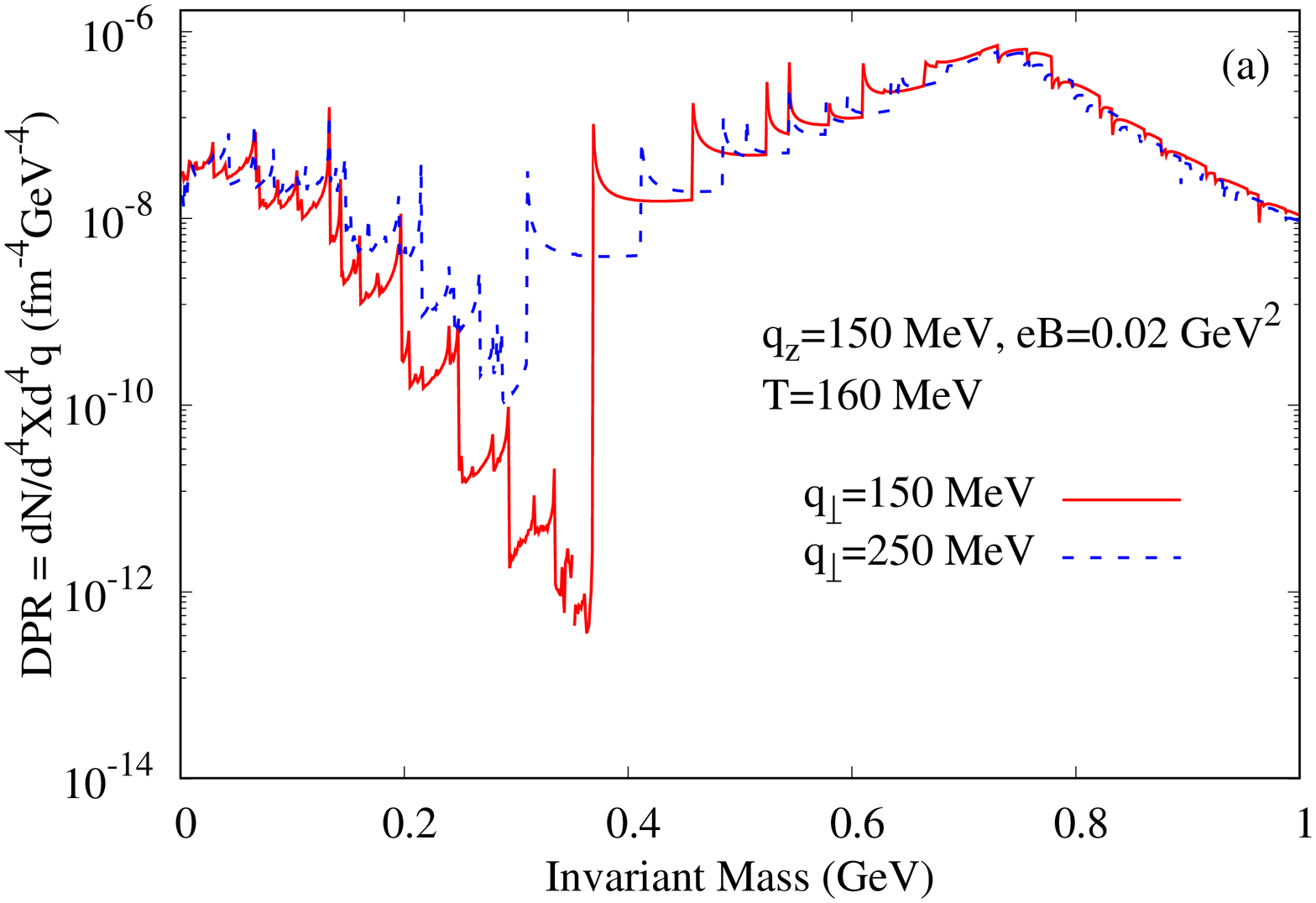}
	\includegraphics[angle = -90, scale=.35]{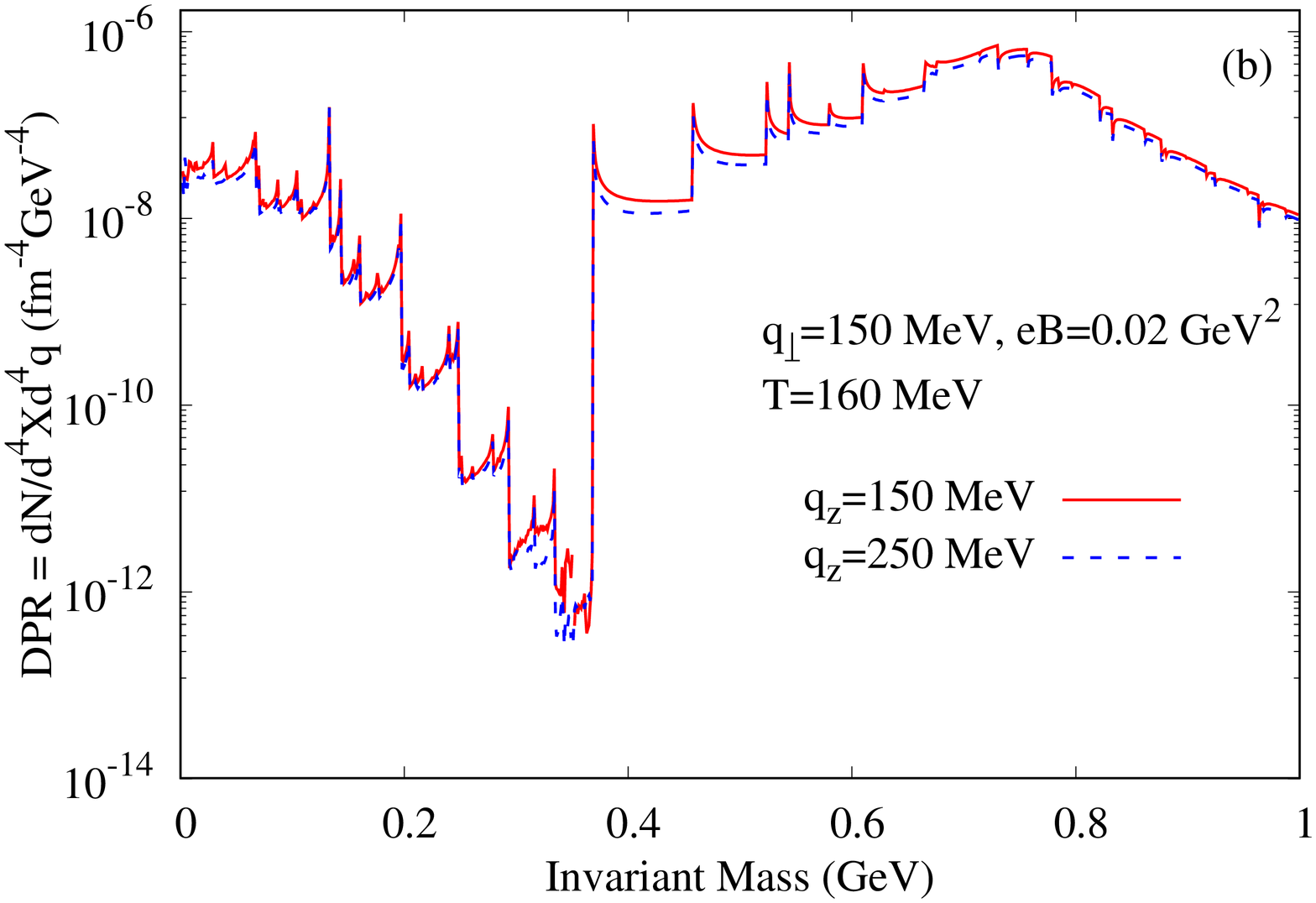}
	\caption{Dilepton production rate (DPR) as a function of the invariant mass at $T=160$ MeV, $eB=0.02$ $\rm GeV^2$ (a) for different values of $q_\perp$ at $q_z= 150$ MeV, (b) for different values of $q_z$ at $q_\perp=150$ MeV,}
	\label{Fig.DPR.qp.qz}
\end{figure}

Now, we turn our attention to the dilepton production rate (DPR) from the hot hadronic matter in the presence of a background magnetic field. In Figs.~\ref{Fig_DPR}(a) and (b), we have  presented  DPR as a function of the invariant mass at $q_\perp$=0.0, $q_z$=150 MeV and temperatures 130 and 160 MeV respectively,  for different values of the magnetic field. The corresponding curves in absence of the background field (grey dotted line) are also shown for comparison which is consistent with the earlier observations by C. Gale and J. Kapusta in Refs.~\cite{Gale:1988vv,Gale:1990pn}. It can be noticed that in both the figures, when the magnetic field is turned on, DPR receives contributions from both Landau cut as well as Unitary-I cut which is understandable from the discussions below Eq.~\eqref{ImPi3} and Eq.~\eqref{Nmununl}. Since the mass of the leptons are much smaller compared to that of $ \pi $-mesons, the threshold invariant mass for dilepton production for all $ eB $ values coincides with Unitary-I cut threshold of {$ \IM~\overline{{D}} $} as evident from Figs.~\ref{Fig_DPR}(a) and (b). Moreover, as we have already justified that both Unitary and Landau cut thresholds are independent of the temperature of the medium, the thresholds of Landau cut contributions which is a purely magnetic field dependent effect, also remain same as observed in Fig.~\ref{Fig_D}(a) for different values of $ eB $. The appearance of non-trivial Landau cut contributions,  leads to significant enhancement in the DPR in the lower invariant mass region which was forbidden in the absence of the background field. Furthermore, at vanishing transverse momentum, for finite values of $ eB $, dilepton production is kinematically forbidden between the Landau and Unitary cut thresholds which can be observed in both the figures Figs.~\ref{Fig_DPR}(a) and (b). The width of this forbidden gap is independent of $ T $ and increases with $ eB $ which can be understood from Eqs.~\eqref{Eq_UCut} and \eqref{Eq_LCut}. Dilepton production considering non-zero values of $q_\perp$ and $q_z$ is presented in the Figs.~\ref{Fig_DPR}(c) and (d). Here, the most interesting observation is that the dilepton production rate becomes continuous and the forbidden gap (existing between Landau cut and Unitary cut when $q_\perp=0$) vanishes. In addition, the DPR is significantly enhanced in the low invariant mass region (Landau cut region). It may be noted that for vanishing $q_\perp$, a pion in Landau level $(n)$ could interact with a pion at Landau levels $(n-1), n, (n+1)$ producing a $\rho^0$-meson. But, there is no such restriction on Landau levels for non-vanishing $q_\perp$, which can be understood by Eq.~\eqref{Nmununl} and discussions below.
The spike-like structures can be seen over the whole range of allowed invariant mass for dilepton production which is a manifestation of the well-known phenomena of `threshold singularities' as discussed earlier. These singularities due to landau level quantization of pions of magnetized hadronic matter can be attributed to the functional dependency of the dilepton production rate as discussed earlier. For given values of the other parameters, we have found that for $eB\ne0$, the overall dilepton production rate is about the same as $eB=0$ at higher invariant mass, i.e, $\sqrt{q^2}>\sqrt{4(m_\pi^2+eB)+q_\perp^2}$. On the other hand, the dilepton production rate is enhanced in low invariant mass region, i.e, $\sqrt{q^2}<\sqrt{4(m_\pi^2+eB)+q_\perp^2}$ (which is absent for $eB=0$) as evident from Figs.~\ref{Fig_DPR}(c) and (d). Finally, it can be inferred that for higher values of temperature, as a consequence of enhancement in the availability of the thermal phase space, the overall magnitude of the DPR increases which is evident from the comparison of Fig.~\ref{Fig_DPR}(a) and (b) or Fig.~\ref{Fig_DPR}(c) and (d).

Figs.~\ref{Fig.DPR.qp.qz} (a) and (b) show dilepton production rate for different values of $q_\perp$ at $q_z=150$ MeV and different values of $q_z$ at $q_\perp=150$ MeV respectively considering  $eB=0.02 \ \rm GeV^2$ and $T=160$ MeV. A similar trend as in Fig.~\ref{Fig_DPR}(d) is observed in both high and low invariant mass region. Moreover, Fig.~\ref{Fig.DPR.qp.qz}(a) shows that, with the increase of the value of $q_\perp$, the Unitary cut threshold shifts towards the lower invariant mass region and the Landau cut threshold shifts towards the higher invariant mass region which is understandable from the discussions below Eqs.~\eqref{ImPi3} and \eqref{Nmununl}. So, there is a combined effect (of Landau and Unitary cut) on the dilepton production rate for the whole range of invariant mass when $q_\perp^2\ge{4(m_\pi^2+eB)}$. On the other hand, \ref{Fig.DPR.qp.qz}(b) shows that DPR decreases with the increase in $q_z$ due to the thermal suppression.
%~~~~~~~~~~~~~~~~~~~~~~~~~~~~~~~~~~~~~~~~~~~~~~~~~~~~~~~~~~~~~~~~~~~~~~~~~~~~~~~~~~~~~~~~~~~~~~~~~~~~~~~~~~~~~~~~~~~~~~~~~~~~~~~~~~~~
\section{Summary \& Conclusion}\label{SC}
In summary, we have presented an analysis of the dilepton production rate from hot hadronic matter under an external magnetic field. We have shown numerical results for  DPR as a function of invariant mass for different values of transverse and longitudinal momenta of the dileptons. The principal component in the DPR is the thermo-magnetic in-medium spectral function of the $\rho^0$ i.e. the imaginary part of the complete interacting $\rho^0$ propagator which has been obtained by solving the Dyson-Schwinger equation containing the one-loop self energy. The self energy of $\rho^0$ in such a thermo-magnetic background is calculated employing the RTF of finite temperature field theory and Schwinger proper-time formalism. The analytic structure is investigated in the complex energy plane; in addition to the usual contribution coming from the Unitary cut beyond the two-pion threshold, we find a non-trivial Landau cut in the physical kinematic region. The appearance of such a non-trivial Landau cut is due to the fact that the charged pions occupy different Landau levels before and after scattering with the $ \rho^0 $ meson which is purely a finite magnetic field effect. Owing to the emergence of the Landau cut, the DPR  yield in the low invariant mass region is non-zero whereas it is absent in the zero field case.  The most interesting finding is the continuous spectrum of DPR owing to shifting of Unitary(Landau) cut thresholds towards lower(higher) values of invariant mass for finite values of $ q_\perp $. However, with vanishing transverse momentum we observe that there exists a forbidden gap between the Landau and Unitary cut thresholds where dilepton production is not  kinematically allowed. The width of the forbidden gap is independent of $ T $ and increases with $ eB $. The enhancement of DPR, in low invariant mass region, is more prominent in case of $q_\perp\ne0$ as compared to $q_\perp=0$ case. This is due to the fact that at $ q_\perp = 0 $ a pion in Landau level $n$ could interact with a pion at Landau levels $(n-1), n, (n+1)$ producing a $\rho^0$-meson, but no such restriction exists for non-vanishing $q_\perp$ resulting in enhanced production in the latter case. Furthermore, with the increase in temperature, the overall magnitude of the DPR is found to increase due to the increase in the availability of the thermal phase space.

 It should be noted that dileptons are produced in all stages of heavy ion collisions. In order to get the dilepton spectrum relevant for experimental observation, one has to integrate the DPR from quark matter as well as hadronic matter over space and time. Although many calculations of dilepton production rate from magnetized quark matter exist in the literature, the emission rate from magnetized hadronic matter evaluated for the first time in this work is an essential contribution to obtain the full spectrum of dileptons from relativistic heavy ion collision.

%~~~~~~~~~~~~~~~~~~~~~~~~~~~~~~~~~~~~~~~~~~~~~~~~~~~~~~~~~~~~~~~~~~~~~~~~~~~~~
\section*{Acknowledgments}
S.G. is funded by the Department of Higher Education, Government of West Bengal, India. 
%~~~~~~~~~~~~~~~~~~~~~~~~~~~~~~~~~~~~~~~~~~~~~~~~~~~~~~~~~~~~~~~~~~~~~~~~~~~~~

%~~~~~~~~~~~~~~~~~~~~~~~~~~~~~~~~~~~~~~~~~~~~~~~~~~~~~~~~~~~~~~~~~~~~~~~~~~~~~~~~~~~~~~~~~~~~~~~~~~~~~~
\appendix 
\section{Comparison with the Expressions of DPR found in the Literature at $B=0$}\label{A1}
Let us change our Cartesian coordinate system to Milne coordinate system via relation 
\begin{eqnarray}
	(q^0,\bm{q}) \equiv (q^0,q_x,q_y,q_z)\to(M_T\cosh y,q_T\cos\phi,q_T\sin\phi,M_T\sinh y)	
\end{eqnarray}
where $q_T=\sqrt{q_x^2+q_y^2}$ is the transverse momentum, 
$M_T=\sqrt{M^2+q_T^2}=\sqrt{q_0^2-q_z^2}$ is the transverse mass, $M=\sqrt{q^2}$ is the invariant mass, and $y=\tanh^{-1}\FB{q_z/q^0}$ is the rapidity. 
Then the infinitesimal four-momentum element $d^4q$ in the Milne system can be written as $d^4q=MdMM_TdM_Td\phi dy$. 
Assuming azimuthal ($\phi$) symmetry, the DPR in Eq.~\eqref{DPR2}, can be integrated to obtain 
\begin{eqnarray}
	\frac{dN}{d^4xdM} &=& \int_{M}^{\infty}M_TdM_T\int_{0}^{2\pi}d\phi\int_{-\infty}^{\infty}dyM\frac{dN}{d^4xd^4q} \nn \\
	&=&\frac{2\alpha^2}{\pi^2M}F_\rho^2m_\rho^2L(M^2)\int_{M}^{\infty}dM_T\int_{-\infty}^{\infty}dy
	\frac{M_T}{e^{(M_T\cosh y)/T}-1} \mathcal{A}. \label{eq.DPR.integrated}
\end{eqnarray}
We now substitute the expression of the spectral function $\mathcal{A}$ from Eq.~\eqref{eq.spec.1} into Eq.~\eqref{eq.DPR.integrated} to obtain
\begin{eqnarray}
	\frac{dN}{d^4xdM} &=& \frac{2\alpha^2}{3\pi^2M}F_\rho^2m_\rho^2L(M^2)\int_{M}^{\infty}dM_T\int_{-\infty}^{\infty}dy \frac{M_T}{e^{(M_T\cosh y)/T}-1} \nn \\ 
	&& \times\TB{\frac{2\text{Im}\Pi_T}{(q^2-m_\rho^2+\text{Re}\Pi_T)^2+(\text{Im}\Pi_T)^2}
		+\frac{\text{Im}\Pi_L}{(q^2-m_\rho^2+\text{Re}\Pi_L)^2+(\text{Im}\Pi_L)^2}}. \label{eq.DPR.integrated.2}
\end{eqnarray}
It has been observed that the difference between the longitudinal and transverse polarization is very small up to reasonably high temperature~\cite{Gale:1990pn,Sarkar:2000ag} 
for the interaction considered here, so that $\text{Im}\Pi_T\approx \text{Im}\Pi_L \approx M\Gamma_\rho(M)$. Also considering approximation $\text{Re}\Pi_{T,L}\approx0$, 
the $dM_Tdy$ integrals of Eq.~\eqref{eq.DPR.integrated.2} can be analytically performed to get the DPR due to pion annihilation following Refs.~\cite{Gale:1988vv,Gale:1990pn,Sarkar:2000ag} as
\begin{eqnarray}
	\frac{dN}{d^4xdM}&=&\frac{4\alpha^2}{\pi^2}\frac{F_\rho^2}{m_\rho^2}MTK_1(M/T)L(M^2)|F_\pi(M)|^2\Gamma_\rho(M)\\
	&=&\frac{\sigma_\pi(M)}{(2\pi)^4}\FB{\frac{F^2_\rho}{4m_\rho^2}g^2_{\rho\pi\pi}}MTK_1(M/T)\FB{1-\frac{4m^2_\pi}{M^2}}
\end{eqnarray}
using the Boltzmann approximation where $K_1$ is the modified Bessel function, $\Gamma_\rho(M)=\frac{g_{\rho\pi\pi}^2}{192\pi}M^5\FB{1-\frac{4m_\pi^2}{M^2}}^{3/2}$ is the 
$\rho^0$-meson decay rate in vacuum~\cite{Mallik:2016anp}, 
\begin{equation}
	|F_\pi(M)|^2=\frac{m_\rho^4}{(q^2-m_\rho^2)^2+\SB{M\Gamma^\rho(M)}^2}
\end{equation}
is the pion form factor~\cite{Bhaduri:1988gc,Ericson:1988gk}, and $\sigma(M)$ is the pion annihilation cross-section given by 
\begin{equation}
	\sigma_\pi(M)=\frac{4\pi\alpha^2}{3M^2}L(M^2)\sqrt{1-\frac{4m_\pi^2}{M^2}}|F_\pi(M)|^2.
\end{equation}

%~~~~~~~~~~~~~~~~~~~~~~~~~~~~~~~~~~~~~~~~~~~~~~~~~~~~~~~~~~~~~~~~~~~~~~~~~~~~~~~~~~~~~~~~~~~~~~~~~~~~~~~~~~~~~~~~~~~~~~~~

\section{$eB$-dependent Vacuum Contribution}\label{Apendx.RePi.VccB}
The expression for $\text{Re}~\Pibar^{\mu\nu}_{\rm~Vac}(q,eB)$ in Eq.~\eqref{eq.repi.B.d3k} is
\begin{equation}
	\Pibar^{\mu\nu}_\text{vac}(q, eB)=i~\sum_{n=0}^{\infty}\sum_{l=0}^{\infty}\int\frac{d^2k_{||}}{\FB{2\pi}^2}\int\frac{d^2k_\perp}{\FB{2\pi}^2}~\frac{\tilde{N}^{\mu\nu}_{nl}\FB{q, k}}{\FB{k^2_{||}-m^2_l+i\epsilon}\FB{\FB{q_{||}+k_{||}}^2-m^2_n+i\epsilon}}
\end{equation}
With $q_\perp=0$, the $\Pibar^{\mu\nu}_\text{vac}(q, eB)$ can be written as~\cite{Ghosh:2019fet}
\begin{eqnarray}
	\Pibar^{\mu\nu}_\text{vac}(q_\parallel, eB)=\Pi^{\mu\nu}_\text{Pure-Vac}(q_\parallel, eB) +\Pi^{\mu\nu}_\text{B-Vac}(q_\parallel, eB) 
\end{eqnarray}
where the explicit form of $\Pi^{\mu\nu}_\text{B-Vac}(q_\parallel, eB)$ is
\begin{eqnarray} \label{Eq_Pi_eB_Vac}
\Pi^\munu_\text{B-Vac}(q_\parallel,B) &=& \frac{-g_{\rho\pi\pi}^2q_\parallel^2}{32\pi^2}\int_{0}^{1}dx
\TB{ \Delta \SB{ \ln\FB{\frac{\Delta}{2eB}}-1 } 
	(q_\parallel^2g^\munu-q_\parallel^\mu q_\parallel^\nu) - 2eB
	\SB{\ln\Gamma\FB{\frac{\Delta}{2eB}+\frac{1}{2}}-\ln\sqrt{2\pi}} (q_\parallel^2g_\parallel^\munu-q_\parallel^\mu q_\parallel^\nu) \right. \nn \\ && \left.
	+ q_\parallel^2\SB{ \Delta+\frac{eB}{2}-\frac{\Delta}{2}
		\SB{\psi\FB{\frac{\Delta}{2eB}+\frac{1}{2}} + \psi\FB{\frac{\Delta}{2eB}+x+\frac{1}{2}  } } }g_\perp^\munu }
\end{eqnarray}
where $\psi(z)$ is the digamma function and $\Delta = m_\pi^2 -x(1-x)q_\parallel^2 - i\epsilon$.
\begin{eqnarray}
	\Delta = m_\pi^2 -x(1-x)q_\parallel^2 - i\epsilon.
\end{eqnarray}
For $q_\perp=0$, the expression ${N}^{\mu\nu}_{nl}$ is found in Eq.~\eqref{Nmununl}. The below results can be obtained from Eq.~\eqref{Nmununl}
\begin{eqnarray}
g_{\mu\nu}{N}^{\mu\nu}_{nl}\FB{q_{\parallel}, k_{\parallel}}&=&\FB{-1}^{n+l}4g^2_{\rho\pi\pi}~\frac{eB}{8\pi} \Big[
\SB{q_{\parallel}^4k_{\parallel}^2+\FB{q_{\parallel}\cdot k_{\parallel}}^2q_{\parallel}^2-2q_{\parallel}^2\FB{q_{\parallel}\cdot k_{\parallel}}^2} \delta^n_l \nn \\ 
&& \hspace{3cm} -\frac{eB}{2}q_{\parallel}^4\SB{\FB{2n+1}\delta^n_l-\FB{n+1}\delta^{n+1}_l-n\delta^n_l}\Big], \label{Eq_Nmumu_B} \\
{N}^{00}_{nl}\FB{q_{\parallel}, k_{\parallel}} &=& \FB{-1}^{n+l}4g^2_{\rho\pi\pi}~\frac{eB}{8\pi}\TB{q_{\parallel}^4k_{0}^2
	+\FB{q_{\parallel}\cdot k_{\parallel}}^2q_{0}^2-2q_{\parallel}^2\FB{q_{\parallel}\cdot k_{\parallel}}q^0k^0}\delta^n_l.\label{Eq_N00_B}
\end{eqnarray}
The corresponding results for zero magnetic field are obtained from Eq.~\eqref{eq.N0} as
\begin{eqnarray}
	g_{\mu\nu}N^{\mu\nu}\FB{q, k}&=&g^2_{\rho\pi\pi}\TB{k^{\mu}k^{\nu}q^4+\FB{q\cdot k}^2q^2-2q^2\FB{q\cdot k}^2}, \\
	N^{00}\FB{q, k}&=&g^2_{\rho\pi\pi}\TB{k_0^2q^4+\FB{q\cdot k}^2q_0^2-2q^2\FB{q\cdot k}q^0k^0}.
\end{eqnarray}

%~~~~~~~~~~~~~~~~~~~~~~~~~~~~~~~~~~~~~~~~~~~~~~~~~~~~~~~~~~~~~~~~~~~~~~~~~~~~~~~~~~~~~~~~~~~~~~~~~~~~~~~~~~~~~~~~~~~~~~~~

\bibliographystyle{apsrev4-1}
\bibliography{Ref}

\end{document}